\documentclass[12pt]{article}

\usepackage[a4paper,left=30mm,right=20mm,top=20mm,bottom=20mm]{geometry}
\usepackage{graphics}
\usepackage{amsmath}
\usepackage{epsfig}
\usepackage{cite}
\usepackage{lineno}

\title{Significance of non-perturbative input to TMD gluon density for hard processes at LHC}

\author{A.A. Grinyuk$^1$, A.V.~Lipatov$^{1,\,2}$, G.I. Lykasov$^1$, N.P.~Zotov$^2$}

\begin{document}

\maketitle

\begin{center}

{\it $^1$Joint Institute for Nuclear Research, Dubna 141980, Moscow Region, Russia}
{\it $^2$Skobeltsyn Institute of Nuclear Physics, Lomonosov Moscow State University, 119991 Moscow, Russia}\\

\end{center}

\vspace{1.0cm}

\begin{center}

{\bf Abstract }

\end{center} 

\indent 
We study the role of the non-perturbative input to the transverse momentum dependent (TMD)
gluon density in hard processes at the LHC. 
We derive the input TMD gluon distribution at a low scale $\mu_0^2 \sim 1$~GeV$^2$
from a fit of inclusive hadron spectra measured at low
transverse momenta in $pp$ collisions at the LHC
and demonstrate that the best description of these spectra 
for larger hadron transverse momenta
can be achieved by matching the derived TMD gluon distribution
with the exact solution of the Balitsky-Fadin-Kuraev-Lipatov (BFKL) 
equation obtained at low $x$ and small gluon transverse momenta outside the saturation region.
Then, we 
extend the input TMD gluon density to higher $\mu^2$ numerically using 
the Catani-Ciafoloni-Fiorani-Marchesini (CCFM) gluon evolution equation.
Special attention 
is paid to phenomenological applications of the obtained
TMD gluon density to some LHC processes, which are sensitive to the gluon
content of a proton.

\vspace{2cm}

\noindent
PACS number(s): 12.38.Bx, 14.65.Dw, 14.65.Fy

\newpage

\section{Introduction} \indent

Numerous experimental studies at the LHC are a challenge to
theoretical QCD motivated approaches and models. In recent years an understanding has been obtained 
that the processes at high
energies and large momentum transfer containing multiple hard
scales require using so-called unintegrated, or transverse momentum dependent
(TMD) parton density functions (PDFs), which have been used within the framework of 
the phenomenological $k_T$-factorization approach\cite{1,2}
for many years. In this approach, the TMD parton
densities are among the main components that determine its
predictive power (see, for example, reviews\cite{3} 
for more information).  The non-perturbative input 
determines behavior of the TMD gluon density 
and production cross sections at a
small gluon transverse momentum $k_T \to 0$
and plays a significant role
in the $k_T$-factorization\cite{4,5,6,7,8,9}.

In our previous papers\cite{10,11} we obtained the non-perturbative input from the
description of the inclusive spectra of hadrons produced in $pp$ collisions at the LHC energies in 
the mid-rapidity region at low 
transverse momenta $k_T\leq 1.5-1.6$~GeV and a starting scale $\mu_0^2 = 1$~GeV$^2$. 
The proposed input is similar to the TMD gluon density 
calculated within the popular color-dipole Golec-Biernat-W\"usthoff (GBW) approach\cite{12}
at large $k_T$ and differs from it at low transverse momenta.
Then, we extended this gluon density to higher $\mu^2$
using the Catani-Ciafoloni-Fiorani-Marchesini (CCFM) evolution equation\cite{13}
and considered deep inelastic $ep$ scattering at HERA. 
We reasonably well described the experimental data on the proton longitudinal structure function $F_L(x,Q^2)$ and 
charm and beauty contribution to the structure function $F_2(x,Q^2)$. 
So, the connection between the soft LHC processes and small $x$ physics at HERA was 
established.

In the present paper we continue our studies\cite{10,11} and 
investigate the role of the non-perturbative input 
to the TMD gluon density in description of hard processes at the LHC. 
We improve the initial TMD gluon distribution
proposed earlier to describe LHC data 
on the inclusive charged hadron spectra at higher transverse momenta
$2.5 < p_T < 4.5$~GeV and numerically extend it to the whole kinematical region 
using the CCFM gluon evolution equation. 
The CCFM equation is the most suitable tool
for our study since it smoothly interpolates between the 
small-$x$ Balitsky-Fadin-Kuraev-Lipatov\cite{14} (BFKL) gluon dynamics
and the conventional   
Dokshitzer-Gribov-Lipatov-Altarelli-Parisi\cite{15} (DGLAP) one.
We extract additional parameters
from a fit to the LHC data on the inclusive $b$-jet production taken by
the CMS and ATLAS Collaborations at high $p_T$ and $\sqrt s = 7$~TeV.
We supply the obtained TMD gluon density with the corresponding
TMD valence and sea quark distributions
calculated in the approximation, where 
the sea quarks occur in the last gluon splitting.
Finally, we discuss several phenomenological applications
of the proposed TMD parton densities to hard LHC processes 
that are most sensitive to the quark and gluon content of the proton.
We use the $k_T$-factorization approach, which is a
commonly recognized tool to investigate hard high-energy processes.
Here we see certain advantages in the fact that, even with the leading-order 
(LO) matrix elements for a hard partonic subprocess, we can take into account a large 
piece of higher order QCD corrections, namely all NLO + NNLO + ... terms containing 
$\log 1/x$ enhancement.

\section{Starting non-perturbative TMD gluon density} \indent

As was mentioned above, the TMD gluon density was obtained\cite{10} within 
the soft QCD model as a 
function of the proton longitudinal momentum fraction $x$ and two-dimensional gluon 
transverse momentum ${\mathbf k}_T$ at a fixed value of the scale $\mu_0^2 = 1$~GeV$^2$. 
It can be 
presented in the simple analytical form:
\begin{equation}
\displaystyle f_g^{(0)}(x,{\mathbf k}_T^2,\mu_0^2) = c_0 c_1 (1-x)^{b} \times \atop {
\displaystyle \times \left[R_0^2(x){\mathbf k}_T^2 + c_2\left(R_0^2(x){\mathbf k}_T^2\right)^{a/2}\right] \exp\left(-R_0(x)
|{\mathbf k}_T|-d\left[R_0^2(x){\mathbf k}_T^2\right]^{3/2}\right)},
\end{equation}

\noindent 
where $R_0^2(x) = (x/x_0)^\lambda/\mu_0^2$, $c_0 = 3 \sigma_0/4\pi^2 \alpha_s$,
$x_0 = 4.21 \cdot 10^{-5}$, $\sigma_0 = 29.12$~mb, $\lambda = 0.22$, and $\alpha_s = 0.2$. 
The parameters $c_1 = 0.3295$, $c_2 = 2.3$, 
$a = 0.7$, $b = 12$ and $d = 0.2$  
were deduced from the best fit of the LHC data on the inclusive spectra of charged
hadrons produced in $pp$ collisions in the mid-rapidity region at low $p_T \leq 1.6$~GeV.
The proposed gluon density differs from the one obtained 
in the GBW model\cite{12} at $|{\mathbf k}_T| < 1$~GeV and 
coincides with the GBW gluon at $|{\mathbf k}_T| > 1.5$~GeV.
Then, it was treated as a starting distribution 
for the CCFM evolution equation and successfully applied to the description of the HERA data
on the proton structure functions $F_L(x,Q^2)$, $F_2^c(x,Q^2)$ and $F_2^b(x,Q^2)$\cite{11}.

However, the proposed non-perturbative gluon density~(1) is not able to describe 
the LHC data on the inclusive spectrum of charged
hadrons at higher transverse momenta $2.5 < p_T < 4.5$~GeV even if
additional pertubative QCD (pQCD) corrections\cite{16,17} are taken into account
at $p_T > 2$~GeV.
Moreover, gluon density~(1) much faster decreases when ${\mathbf k}_T^2$ grows 
compared to the solution of the BFKL equation outside of the saturation region\cite{14,15,18}.
Therefore, we modify the gluon density given by~(1) at $|{\mathbf k}_T| > 2 - 3$~GeV 
to describe the LHC data on the charged hadron production at $2.5 < p_T < 4.5$~GeV. 
Then we
match it with the TMD gluon obtained in\cite{18} as the solution of the linear BFKL equation
at low $x$, which results in flatter ${\mathbf k}_T^2$ 
behavior.
The modified starting TMD gluon density can be presented in the following form:
\begin{equation}
\displaystyle f_g^{(0)}(x,{\mathbf k}_T^2,\mu_0^2) = c_0 c_1 (1-x)^{b} \times \atop {
\displaystyle \times \left[R_0^2(x){\mathbf k}_T^2 + c_2\left(R_0^2(x){\mathbf k}_T^2\right)^{a/2}\right] \exp\left(-R_0^2(x) {\mathbf k}_T^2 -d\left[R_0^2(x){\mathbf k}_T^2\right]^{3/2}\right) + \atop {
\displaystyle + \, c_0 \left({x\over x_0}\right)^n \exp \left[- k_0^2 {R_0(x)\over |{\mathbf k}_T|}\right] f_g(x,{\mathbf k}_T^2), }} 
\end{equation}

\noindent 
where $k_0 = 1$~GeV, $\mu_0^2 = 1.1$~GeV$^2$, and 
$n \simeq 0.81$. The function $f_g(x,{\mathbf k}_T^2)$
obeying the BFKL equation at not very large ${\mathbf k}_T^2$ reads\cite{18}
\begin{equation}
f_g(x,{\mathbf k}_T^2) = \alpha_s^2 \, x^{-\Delta} \, t^{-1/2} {1 \over v} \exp \left[ - { \pi \ln^2 v \over t}\right],
\end{equation}

\noindent 
where $t = 14\,\alpha_s N_c \, \zeta(3) \ln(1/x)$, $v = |{\mathbf k}_T|/\Lambda_{\rm QCD}$,
and $\Delta = 4 \, \alpha_s N_c \ln 2 /\pi$.
It is important that the third term in~(2) is only non-zero
at $|{\mathbf k}_T| \ll \Lambda_{\rm QCD} \, (1/x)^{\delta}$
with $\delta = \alpha_s N_c$.
The details of the calculation and the relation 
between the TMD gluon density and the inclusive hadron spectra $\rho_h(y\simeq 0,p_T)\equiv E \, d^3\sigma/d^3 p$ 
are given in
our previous papers\cite{10,11}.
These spectra are presented as a sum of two parts\cite{16,17}:
\begin{equation}
  \rho_h(y\simeq 0,p_T) = \rho_q(y\simeq 0,p_T) + \rho_g(y\simeq 0,p_T),
\end{equation}

\noindent
where $\rho_q$ is the quark contribution calculated within the 
quark-gluon string model (QGSM)\cite{19,20,21} and $\rho_g$ is the gluon contribution, 
which can be calculated using the proposed TMD gluon density~(2) and~(3).
Taking into account the energy $\sqrt s$ dependence of the $\rho_q$ and $\rho_g$ parts, 
the spectrum $\rho_h(y\simeq 0,p_T)$ can be written in the following 
form:
\begin{equation}
  \rho_h(y\simeq 0,p_T) = \left[\phi_q(y\simeq 0,p_T) + \phi_g(y\simeq 0,p_T) \left(1 - {\sigma_{nd} \over g (s/s_0)^{\Delta}}\right)\right] g (s/s_0)^{\Delta},
\end{equation} 

\noindent
where $g = 21$~mb, $\Delta = \alpha_P(0) - 1 \simeq 0.12$, 
$\alpha_P(0)$ is the Pomeron intercept and $\sigma_{nd}$
is the non-diffractive cross section given by the sum of 
$n$ pomeron chain production cross sections. The 
functions $\phi_q(y\simeq 0,p_T)$ and $\phi_g(y\simeq 0,p_T)$
are evaluated in\cite{16,17} and can be 
presented as
\begin{equation}
\phi_q(y\simeq 0,p_T)=A_q \exp \left(-p_T/C_q\right),
\end{equation} 
\begin{equation}
\phi_g(y\simeq 0,p_T)=A_g \sqrt{p_T} \exp \left(-p_T/C_g\right),
\end{equation} 

\noindent
where the parameters $A_q = 3.68$~GeV$^{-2}$, $A_g = 1.7249$~GeV$^{-2}$, $C_q = 0.147$~GeV, and
$C_g= 0.289$~GeV were obtained from the combined
fit of the NA61\cite{22} and LHC\cite{23} data taken at
different energies $\sqrt s$\cite{16}.
These parameters, of course, differ from the ones obtained earlier\cite{10}
due to another form of TMD gluon density~(2) and~(3) used in the fit procedure.
The latter leads, in addition, to the values of the parameters $d = 0$ and $a = 0.3$ in~(2).
In this way the LHC data\cite{23} are 
fitted with $\chi^2/n.d.f = 0.998$.
The parameters $b = 6.57$ and $\alpha_s = 0.18$ were obtained from the
best fit of the CMS data on the inclusive $b$-jet production at $|y| < 0.5$ (see below).
Let us stress here that since the parameters of nonperturbative input (2) --- (3) 
were obtained from the description of the LHC and NA61 data\cite{22,23}, possible 
higher-order corrections (see, for example,\cite{24,25,26}) 
to the leading-order BFKL motivated ${\mathbf k}_T$-dependence of the proposed 
gluon input at low $x$ (as well as saturation dynamics) are effectively included.

The inclusive spectra of $\pi^-$ mesons produced in 
$pp$ collisions at the initial momenta $31$ and $158$~GeV 
are presented in Fig.~1 as a function of the transverse mass $m_T = \sqrt{m_\pi^2 + p_T^2}$.
Using the first part of the spectrum $\phi_q(y\simeq 0, p_{T})$, connected with
the quark contribution in the conventional string model\cite{19}, one can reasonably describe 
the NA61 data\cite{22} at low $m_T < 1$~GeV. 
The inclusion of the second part of the spectrum, which is due to the gluon contribution, 
allows us to describe the NA61 data up to $m_T \sim 1.5$~GeV.
A similar description of the experimental data on the inclusive spectra of charged 
hadrons (mainly pions and/or kaons) is achieved at the LHC (see Fig.~2).
In addition to the soft part,
we include the pQCD corrections\cite{16,17}. 
The latter, made at the leading order (LO), are divergent at low  
transverse momenta. Therefore, the kinematical region $p_T \sim 1.8 - 2.2$~GeV 
can be treated as the matching 
region of the non-pertubative QCD (soft QCD) and pQCD calculations.
One can see that the inclusive hadron spectra at the LHC 
can be well described in a wide 
region of transverse momenta by matching these two approaches,
and the proposed input TMD gluon density~(2) and (3) plays a crucial 
role in the description of these data at low hadron transverse momenta.

\section{CCFM-evolved TMD parton densities} \indent

The average gluon transverse momentum $\langle |{\mathbf k}_T| \rangle$,
generated by the TMD gluon distribution defined above
is $\langle |{\mathbf k}_T| \rangle \sim 1.9$~GeV at
$10^{-7} < x < 1$, which is close to the non-perturbative 
QCD regime.
Therefore, we can treat the proposed TMD gluon density 
as a starting one and apply the CCFM  
equation to extend it to the whole kinematical region.
The CCFM evolution equation resums large logarithms $\alpha_s^n \ln^n 1/(1-x)$ 
in addition to $\alpha_s^n \ln^n 1/x$ ones and 
introduces angular ordering of initial  
emissions to correctly treat gluon coherence effects.
In the limit of asymptotic energies, it is almost 
equivalent to BFKL, but also similar to the
DGLAP evolution for large $x$ and high $\mu^2$\cite{13}.

In the leading logarthmic approximation, the CCFM equation with 
respect to the evolution (factorization) 
scale $\mu^2$ can be written as\cite{13}
\begin{equation}
\displaystyle f_g(x,{\mathbf k}_T^2,\mu^2) = f_g^{(0)}(x,{\mathbf k}_T^2,\mu_0^2) \Delta_s(\mu^2,\mu_0^2) + \atop { 
\displaystyle + \int {dz\over z} \int {d q^2\over q^2} \theta(\mu - z q) \Delta_s(\mu^2,z^2 q^2)} P_{gg}(z,q^2,{\mathbf k}_T^2) 
f_g(x/z,{\mathbf k^\prime}_T^2,q^2),
\end{equation}

\noindent 
where ${\mathbf k^\prime}_T = {\mathbf q} (1 - z) + {\mathbf k}_T$ and the 
Sudakov form factor $\Delta_s(q_1^2,q_2^2)$ describes the probability of no radiation
between $q_2^2$ and $q_1^2$. 
The first term in the CCFM equation~(8), which is initial TMD gluon density 
multiplied by the Sudakov form factor $\Delta_s(\mu^2,\mu^2_0)$, makes the contribution of
non-resolvable branchings between the starting scale $\mu_0^2$ and the factorization scale
$\mu^2$, the second term describes the details of the QCD evolution expressed by the 
convolution of the CCFM splitting function $P_{gg}(z,q^2,{\mathbf k}_T^2)$ with the
gluon density $f_g(x,{\mathbf k}_T^2,\mu^2)$ and the Sudakov form factor $\Delta_s(\mu^2,q^2)$, 
and the theta function introduces the angular ordering condition. 
The evolution scale $\mu^2$ is defined by the maximum allowed angle for any gluon 
emission\cite{13}.

The CCFM equation~(8) describes only
the emission of gluons, while quark emissions are left aside.
In order to calculate the TMD valence quark denisities, we have to replace in~(8)
the gluon splitting function $P_{gg}(z,q^2,{\mathbf k}_T^2)$ 
by the quark one $P_{qq}(z,q^2,{\mathbf k}_T^2)$ \cite{27,28}. 
The starting TMD valence quark distribution can
be parameterized using standard collinear PDFs $xq_v(x,\mu^2)$ as
\begin{equation}
  f_{q_v}^{(0)}(x,{\mathbf k}_T^2,\mu_0^2) = xq_v(x,\mu_0^2) \, {2\over \mu_0^2} \exp \left[ - { {\mathbf k}_T^2 \over \mu_0^2/2} \right].
\end{equation}

\noindent
Numerically, we applied the LO  
parton densities from the MSTW'2008 set\cite{29}.
The exact analytical expressions for the splitting functions $P_{gg}(z,q^2,{\mathbf k}_T^2)$,
$P_{qq}(z,q^2,{\mathbf k}_T^2)$,
and Sudakov form factor can be found, for example, in\cite{30}.
Concerning the TMD sea quark density, we calculate it using the approximation where the sea quarks 
occur in the last gluon-to-quark splitting. 
At the next-to-leading logarithmic accuracy $\alpha_s (\alpha_s \ln x)^n$ the TMD sea 
quark distribution
can be written\cite{7} as:
\begin{equation}
  f_{q_s}(x,{\mathbf k}_T^2,\mu^2) = \int \limits_x^1 {dz \over z} \int d{\mathbf q}_T^2
    {1\over {\mathbf \Delta}^2} {\alpha_s \over 2\pi} P_{qg}(z,{\mathbf q}_T^2,{\mathbf \Delta}^2) f_g(x/z,{\mathbf q}_T^2, \bar \mu^2),
\end{equation}

\noindent
where $z$ is the fraction of the gluon light cone momentum carried out by
the quark, and $\mathbf \Delta = {\mathbf k}_T - z{\mathbf q}_T$. 
The sea quark evolution is driven by the off-shell gluon-to-quark
splitting function $P_{qg}(z,{\mathbf q}_T^2,{\mathbf \Delta}^2)$\cite{31}:
\begin{equation}
  P_{qg}(z,{\mathbf q}_T^2,{\mathbf \Delta}^2) = T_R \left({\mathbf \Delta}^2\over {\mathbf \Delta}^2 + z(1-z)\,{\mathbf q}_T^2\right)^2
    \left[(1 - z)^2 + z^2 + 4z^2(1 - z)^2 {{\mathbf q}_T^2\over {\mathbf \Delta}^2} \right],
\end{equation}

\noindent 
where $T_R = 1/2$. The splitting function $P_{qg}(z,{\mathbf q}_T^2,{\mathbf \Delta}^2)$
was obtained by generalizing to finite transverse momenta of 
the two-particle irreducible kernel expansion\cite{32}.
It takes into account the small-$x$ enhanced transverse momentum dependence 
up to all orders in the strong coupling constant, and reduces to the conventional splitting
function at the lowest order for $|\mathbf q_T| \to 0$.
The scale $\bar \mu^2$ was defined\cite{33} from the angular ordering condition which is natural
from the point of view of the CCFM evolution: $\bar \mu^2 = {\mathbf \Delta}^2/(1-z)^2 + {\mathbf q}_T^2/(1-z)$.

The CCFM evolution equation with the starting TMD gluon and quark distributions 
given by~(2) --- (3) and (9)  
was solved numerically\footnote{Authors are very grateful to Hannes Jung for providing 
us with the appropriate numerical code.} in the leading logarithmic approximation
using the \textsc{updfevolv} routine\cite{30}.
Thus, the TMD gluon and valence quark densities were 
obtained for any $x$, ${\mathbf k}_T^2$, and $\mu^2$ values.
The TMD sea quark distributions can be evaluated according to~(10) and~(11). 

The gluon density $f_g(x,{\mathbf k}_T^2,\mu^2)$
obtained according to~(2),~(3) and~(8),
labeled below as {\it Moscow-Dubna 2015}, or MD'2015,
is shown in Fig.~3 as a 
function of ${\mathbf k}_T^2$ for different values of $x$ and $\mu^2$.
Additionally, we plot the TMD gluon distribution\cite{34} (namely, the set A0)
which is widely discussed in the literature and commonly used in the applications. 
One can observe some difference in the absolute normalization and shape between 
both TMD gluon distributions. Below we will consider the corresponding phenomenological 
consequences for several LHC processes.

\section{Phenomenological applications} \indent

We are now in a position to apply the proposed TMD parton densities
to some processes studied at hadron colliders.
In the present paper we consider the inclusive production
of $b$-jets, $B^+$ and $D^*$ mesons as well as the associated production of $W^\pm$ or $Z/\gamma^*$ bosons
and hadronic jets at the LHC conditions. We also study the charm and beauty contribution to the 
proton structure function $F_2(x,Q^2)$ and the longitudinal proton structure function $F_L(x,Q^2)$.
These processes are known to be strongly sensitive to the gluon and/or quark content of
the proton.

In according to the $k_T$-factorization prescription\cite{1,2},
the cross sections of the processes under consideration can be written as
\begin{equation}
  \displaystyle \sigma = \int dx_1 dx_2 \int d {\mathbf k}_{1T}^2 d {\mathbf k}_{2T}^2 \, f_{q/g}(x_1,{\mathbf k}_{1T}^2,\mu^2) f_{q/g}(x_2,{\mathbf k}_{2T}^2,\mu^2) \times \atop {
  \displaystyle  \times d\hat \sigma(x_1, x_2, {\mathbf k}_{1T}^2,{\mathbf k}_{2T}^2, \mu^2) },
\end{equation}

\noindent 
where $\hat \sigma(x_1, x_2, {\mathbf k}_{1T}^2,{\mathbf k}_{2T}^2, \mu^2)$ are relevant 
off-shell (depending on the transverse momenta of incoming particles) 
partonic cross sections. The detailed description of the 
calculation steps (including the evaluation of the off-shell 
amplitudes) can be found in our previous papers\cite{35,36,37,38}.
Here we only specify the essential numerical parameters.
Following\cite{39}, we set the charmed and beauty quark masses $m_c = 1.4$~GeV and 
$m_b = 4.75$~GeV, 
$D^*$ and $B^+$ meson masses $m_{D^*} = 2.01$~GeV and $m_{B^+} = 5.28$~GeV, masses of
gauge bosons $m_W = 80.403$~GeV and $m_Z = 91.1876$~GeV, 
$Z$ boson decay width $\Gamma_Z = 2.4952$~GeV, and
Weinberg mixing angle $\sin^2 \theta_W = 0.231$. We use the two-loop formula for 
the strong coupling constant (as it is implemented in the \textsc{updfevolv} routine) 
with $n_f= 4$ active quark flavors at
$\Lambda_{\rm QCD} = 200$~MeV and apply the running QCD and QED coupling constants.
We use the factorization and renormalization scales $\mu_F$ and $\mu_R$
according to the process under consideration\cite{35,36,37,38}. Additionally, we 
estimate the theoretical uncertainty coming from the
renormalization scale.

The multidimensional integration was performed
by the Monte Carlo technique, using the routine \textsc{vegas}\cite{40}.
The corresponding C++ code is available from the authors on request.

\subsection{Proton structure functions $F_2^c$, $F_2^b$ and $F_L$} \indent

The charm and beauty contributions to the proton structure 
function $F_2(x,Q^2)$ and the longitudinal structure function $F_L(x,Q^2)$
are directly connected to the gluon content of the
proton. The structure function $F_L(x,Q^2)$ is equal to zero in the parton model 
with spin $1/2$ partons and only has nonzero values within the pQCD.
The experimental data on these structure functions were obtained\cite{41,42,43,44,45}
by the H1 and ZEUS Collaborations at HERA.
Our consideration is based on the formalism\cite{35},
and here we present the main results only. 

The results of our calculations are presented in Figs.~4 --- 6.
The solid curves correspond to the predictions obtained with the 
MD'2015 gluon distribution, the upper and lower dashed curves represent 
the estimate of the corresponding theoretical uncertainties (as given by the usual 
scale variations) and the dash-dotted curves correspond to the results
obtained with the A0 gluon density.
Additionally, we plot the data/theory ratios for the longitudinal structure function $F_L(x,Q^2)$.
One can see that the predictions corresponding to the proposed MD'2015 
gluon distribution agree well with the H1 
and ZEUS data in the whole kinematical region of $x$ and $Q^2$
within the theoretical and experimental uncertainties.
It only tends to slightly overestimate the $F_L$ data\cite{41} at very low $Q^2$ 
but still agrees with them within the theoretical and experimental uncertainties.
Therefore, the main conclusion of\cite{10,11}, where the link 
between soft processes at the LHC and low-$x$ physics at HERA was pointed out, is confirmed.
At the same time, the A0 gluon density 
does not reproduce the
shape of the structure functions $F_2^c$ and $F_2^b$ at low $Q^2$.
It means that the influence of the shape and other parameters of the initial 
non-perturbative gluon distribution on the description of experimental 
data is significant for a wide region of $x$ and $Q^2$.
We conclude that the best description of the HERA data\cite{41,42,43,44,45} is 
achieved with the MD'2015 gluon density.

\subsection{Inclusive $b$-jet production}\indent

As is well known, beauty quarks at the LHC energies are produced mainly via 
standard QCD gluon-gluon fusion subprocess; therefore, 
the corresponding total and differential cross sections
are strongly sensitive to the gluon content of the proton.
In the $k_T$-factorization approach this was demonstrated
in our previous paper\cite{36}.
Below we probe the MD'2015 gluon density in the
inclusive $b$-jet production at the LHC.
The CMS Collaboration measured the 
$b$-jet cross sections in five $b$-jet rapidity regions, namely,
$|y| < 0.5$, $0.5 < |y| < 1$, $1 < |y| < 1.5$, $1.5 < |y| < 2$,
and $2 < |y| < 2.2$ as a function of the jet transverse momentum at 
$\sqrt s = 7$~TeV\cite{46}.
The ATLAS Collaboration performed the measurements
at central rapidities $|y| < 2.1$\cite{47}. 

Our main results are presented in Figs.~7 and~8,
where we plot the calculated transverse momentum distributions
of $b$-jets compared to the LHC data as well as the corresponding data/theory ratios.
We obtained a good description of the data using the MD'2015 gluon
distribution. The shape and absolute normalization of the measured $b$-jet
cross sections are reproduced well. 
Moreover, the differential cross section as a function of
the angular separation $\Delta \phi$ between 
two $b$-jets measured at $p_T > 40$~GeV and $M > 110$~GeV,
where $M$ is the invariant mass of the produced $b\bar b$ pair,
is also well described. This observable is known to be very 
sensitive to the ${\mathbf k}_T^2$-behavior of the TMD gluon 
distribution.
The predictions based 
on the A0 gluon density lie below the 
data at high $p_T > 40$ --- $50$~GeV.
Once again, we conclude that the best description of the LHC data\cite{46,47} 
is achieved with the proposed MD'2015 gluon density.

\subsection{$B^+$ meson production}\indent

Besides the $b$-jet production, we probe the proposed MD'2015 gluon density in the
inclusive $B^+$ meson production at the LHC, which was measured
by the CMS\cite{48}, ATLAS\cite{49}, and LHCb\cite{50} Collaborations.
Our basic formulas and the description of the calculation details
are collected in\cite{36,37}.
Here we only note that
we convert beauty quarks produced in the hard subprocess 
into $B^+$ mesons using the Peterson
fragmentation function\footnote{Of course, the predicted transverse momentum distributions are 
sensitive to the quark-to-hadron fragmentation function.
This dependence was studied earlier\cite{36} and not
considered in the present paper.}
with the usual shape parameter $\epsilon_b = 0.006$\cite{51}.
Following\cite{39}, we set the branching fraction $f(b \to B^+) = 0.398$.
The CMS Collaboration reported the total and differential $B^+$ 
cross sections measured at central rapidities $|y| < 2.4$ for $p_T > 5$~GeV and 
$\sqrt s = 7$~TeV\cite{48}. 
The ATLAS Collaboration presented $B^+$ meson cross sections
for $|y| < 2.25$ and $9 < p_T < 120$~GeV\cite{49}.
The LHCb Collaboration measured $B^\pm$ meson cross sections
in the forward rapidity region $2 < y < 4.5$ at $p_T < 40$~GeV\cite{50}.

Our main results are presented in Figs.~9 --- 11 compared with the LHC data
and corresponding data/theory ratios.
One can see that a good description of the $B^+$ meson
transverse momentum and rapidity distributions
is achieved using both TMD gluon densities under consideration. 
The results obtained with the MD'2015 gluon density 
lie somewhat above the A0 ones and tend to slightly overestimate the measured
cross sections at high transverse momenta but agree with the 
data within the uncertainties. 
However, both considered TMD gluon densities give similar behavior
of the rapidity distributions.
We would like to point out the remarkable
description of the LHCb data in the forward
rapidity region that extends essentially the applicability area
of the proposed MD'2015 gluon distribution.

\subsection{$D^*$ meson production}\indent

Similar to beauty quarks, charmed quarks 
are mainly produced at the LHC via the gluon-gluon fusion subprocess, and therefore,
the charm production cross section is also sensitive to the gluon density function.
The transverse momentum distributions of several charmed mesons
($D^*$, $D^\pm$, $D^0$, $D_s$) were measured by the LHCb Collaboration at 
forward rapidities, $2 < y < 4.5$\cite{52}.
As a representative example, below we consider $D^*$ meson production. 
To 
convert $c$-quarks into $D^*$ mesons, we apply the
non-perturbative
fragmentation function\cite{53,54,55},
which is often used in the collinear QCD calculations.
We set the branching fraction $f(c \to D^*) = 0.255$\cite{39}.

Our numerical results are presented in Fig.~12.
We find reasonably good agreement between 
our predictions obtained using the MD'2015 gluon distribution and the LHCb data\cite{52},
which demonstrates again a wide area of its applicability.
The A0 gluon density predicts flatter $D^*$ transverse momentum
distributions and does not contradict the data.

\subsection{Associated $W^\pm +$~jet and $Z/\gamma^* +$~jet production}\indent

Contrary to the processes considered above, where only the
TMD gluon distribution was probed, the
associated production of gauge ($W^\pm$ or $Z$) bosons and hadronic jets at the LHC
offers high sensitivity to both quark and gluon density functions in the proton.
The experimental data on the $W^\pm +$~jet and $Z/\gamma^* +$~jet 
production were obtained
by the ATLAS Collaboration\cite{56,57}. 
Below we apply the TMD quark and gluon densities
from the MD'2015 set to describe the LHC data. Our consideration 
is based on the off-shell amplitudes of
the quark-gluon scattering subprocesses
$q g^* \to W^\pm q^\prime$ and 
$q g^* \to Z/\gamma^* q$
which include the subsequent decays $W^\pm\to l^\pm\nu$ and $Z/\gamma^* \to l^+l^-$ 
derived in\cite{38}. Other details of the calculations can be found there.

The ATLAS data\cite{56} on the associated $W^\pm$ and jet production refer
to the kinematical region defined as 
$p_T^l > 25$~GeV, $|\eta^l| < 2.5$, 
$E_T^{\rm miss} > 25$~GeV, $m_T(W) > 40$~GeV, 
$p_T^{\rm jet} > 30$~GeV, and $|y^{\rm jet}| < 4.4$,
where $m_T(W)$ is the transverse mass of the produced $W^\pm$ boson,
$\eta^l$ and $p_T^l$ are the decay lepton pseudo-rapidity and transverse
momentum, $y^{\rm jet}$ and $p_T^{\rm jet}$ are the rapidity and transverse 
momentum of the final hadronic jet, respectively, and $E_T^{\rm miss}$ is the
missing transverse energy. The measurements of the $Z/\gamma^* +$~jet 
production were performed\cite{57} at $66 < M < 116$~GeV, $p_T^l > 20$~GeV, 
$|\eta^l| < 2.5$, $p_T^{\rm jet} > 20$~GeV, and $|y^{\rm jet}| < 4.4$,
where $M$ is the invariant mass of the produced lepton pair.
Our main numerical results are shown in Fig.~13 in comparison 
with the ATLAS data\cite{56,57}.
Additionally, we plot the corresponding data/theory ratios.
We obtained a good description of these data with both
TMD parton densities.
The latter, in particular, demonstrates that 
the TMD quark and gluon distributions
from the MD'2015 set
are reliable at relatively large scales, up to $\mu^2 \sim m_Z^2$. 
However, we note that the full hadron-level Monte-Carlo generator 
\textsc{cascade}\cite{58}, which 
uses the CCFM evolution equation for the initial state gluon emissions,
is needed for a more detailed analysis
of the associated production of gauge bosons and hadronic jets at the LHC.
It is connected with a more accurate jet selection algorithm 
implemented in \textsc{cascade}, as was explained previously\cite{38}.

\section{Conclusion} \indent

We fitted the experimental data on the inclusive spectra of the charged particles 
produced in the central $pp$ collisions at the LHC to
determine the TMD gluon density in a proton at the starting scale 
$\mu_0^2 \sim 1$~GeV$^2$. We demonstrated that the best 
description of these spectra can be achieved 
by matching the derived TMD gluon distribution
with the exact solution of the Balitsky-Fadin-Kuraev-Lipatov (BFKL) 
equation obtained at low $x$ and small gluon transverse momenta outside
the saturation region.
Moreover, we established that the parameters of this fit did not depend on the
initial energy in a wide energy interval.
The average gluon transverse momentum generated by this modified TMD gluon density 
is about $1.9$~GeV in a wide region of $x$ and is 
close to the non-perturbative QCD regime.
Then, we extended the derived TMD gluon density to higher $\mu^2$ using numerical
solution of the CCFM gluon evolution equation.
Additionally, 
we supplied the calculated TMD gluon density with the
TMD valence and sea quark distributions. The latter was evaluated
in the approximation where the gluon-to-quark splitting occurred at the last evolution step
using the TMD gluon-to-quark splitting function. This function contains 
all single logarithmic small-$x$ corrections in any order of perturbation theory.

Special attention was paid to the phenomenological applications
of the proposed MD'2015 parton distributions to the hard processes.
We considered the inclusive production
of $b$-jets, $B^+$ and $D^*$ mesons and the
associated production of $W^\pm$ or $Z/\gamma^*$ bosons
and hadronic jets at the LHC energies,
and also the charm and beauty contribution to the 
proton structure function $F_2(x,Q^2)$ and the longitudinal proton structure function $F_L(x,Q^2)$. 
We demonstrated significant influence of the initial 
non-perturbative gluon distribution on the description of experimental 
data. We showed that the LHC data could be well described 
using the MD'2015. 

\section{Acknowledgements} \indent

We thank H.~Jung for his extreme help in the calculation of the CCFM evolution and very 
useful discussions and comments. 
We also thank L.N.~Lipatov and B.I.~Ermolaev for helpful discussion 
and M.A.~Malyshev for careful reading of the manuscript. 
This research was supported by the FASI of Russian Federation
(grant NS-3042.2014.2).
A.V.L. and N.P.Z. are also grateful to the DESY Directorate for the
support within the framework of the Moscow --- DESY project on Monte-Carlo implementation for
HERA --- LHC.

\newpage

\begin{figure}
\begin{center}
\epsfig{figure=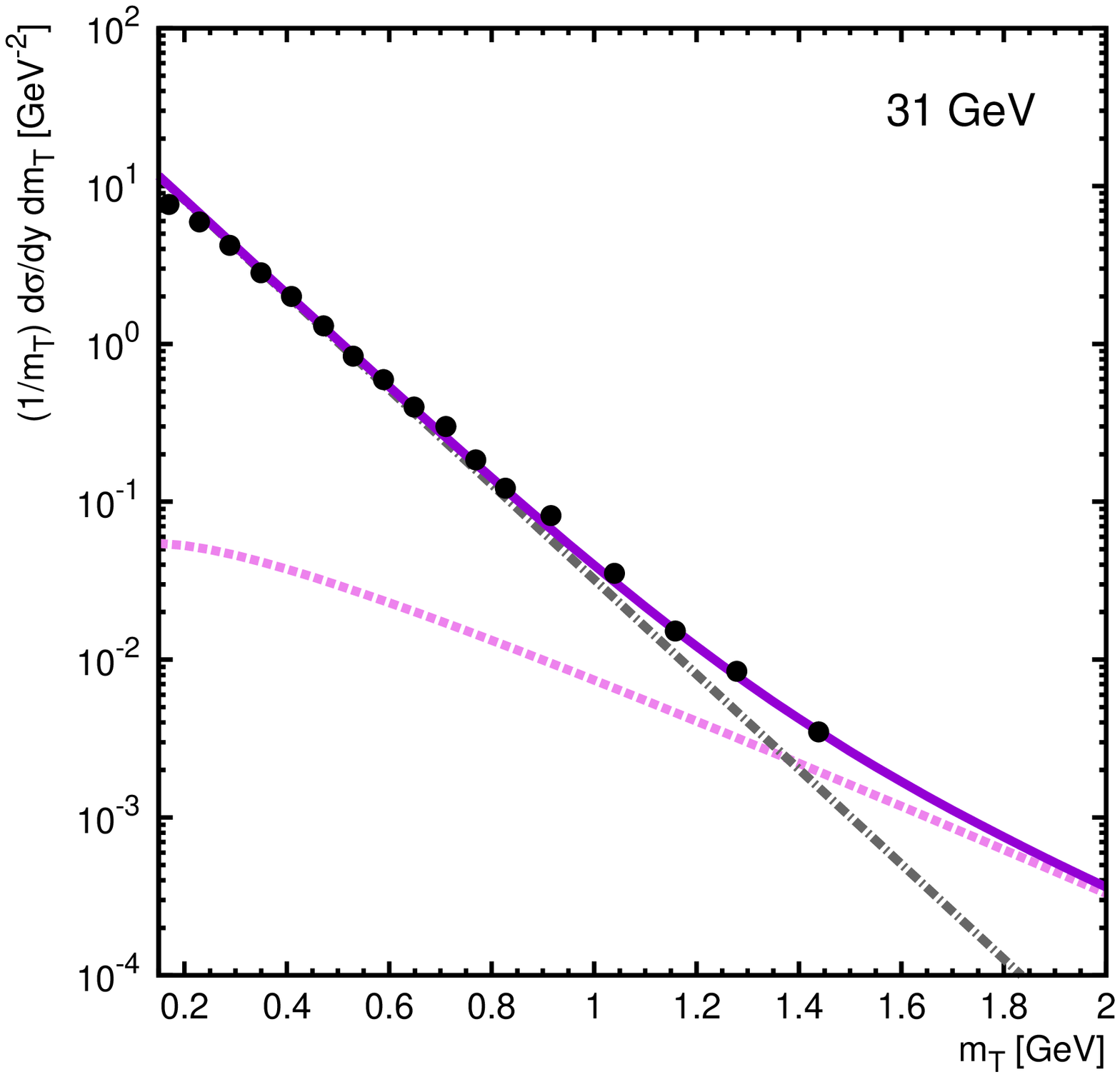, width = 5.25cm, height = 8.0cm, angle = 270}
\vspace{0.7cm} \hspace{-1.1cm}
\epsfig{figure=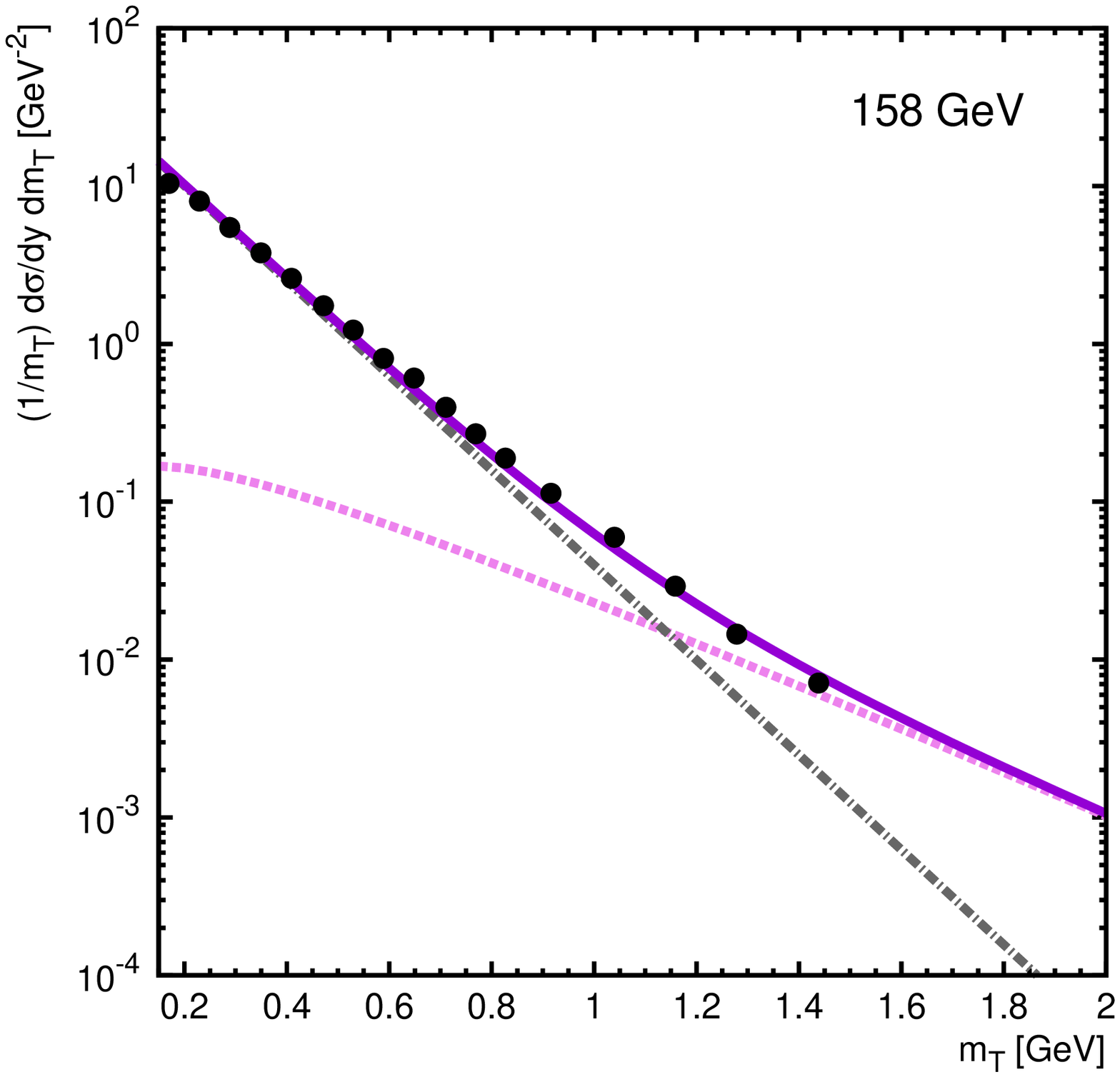, width = 5.25cm, height = 8.0cm, angle = 270}
\caption{The inclusive cross sections of the $\pi^-$ meson production in 
the $pp$ collisions at the initial momenta $31$ and $158$~GeV
as a function of the transverse mass.
The dashed and dash-dotted curves correspond to the gluon and quark contributions, respectively.
The solid curves represent their sum.
The experimental data are from NA61\cite{22}.}
\label{fig1}
\end{center}
\end{figure}

\begin{figure}
\begin{center}
\epsfig{figure=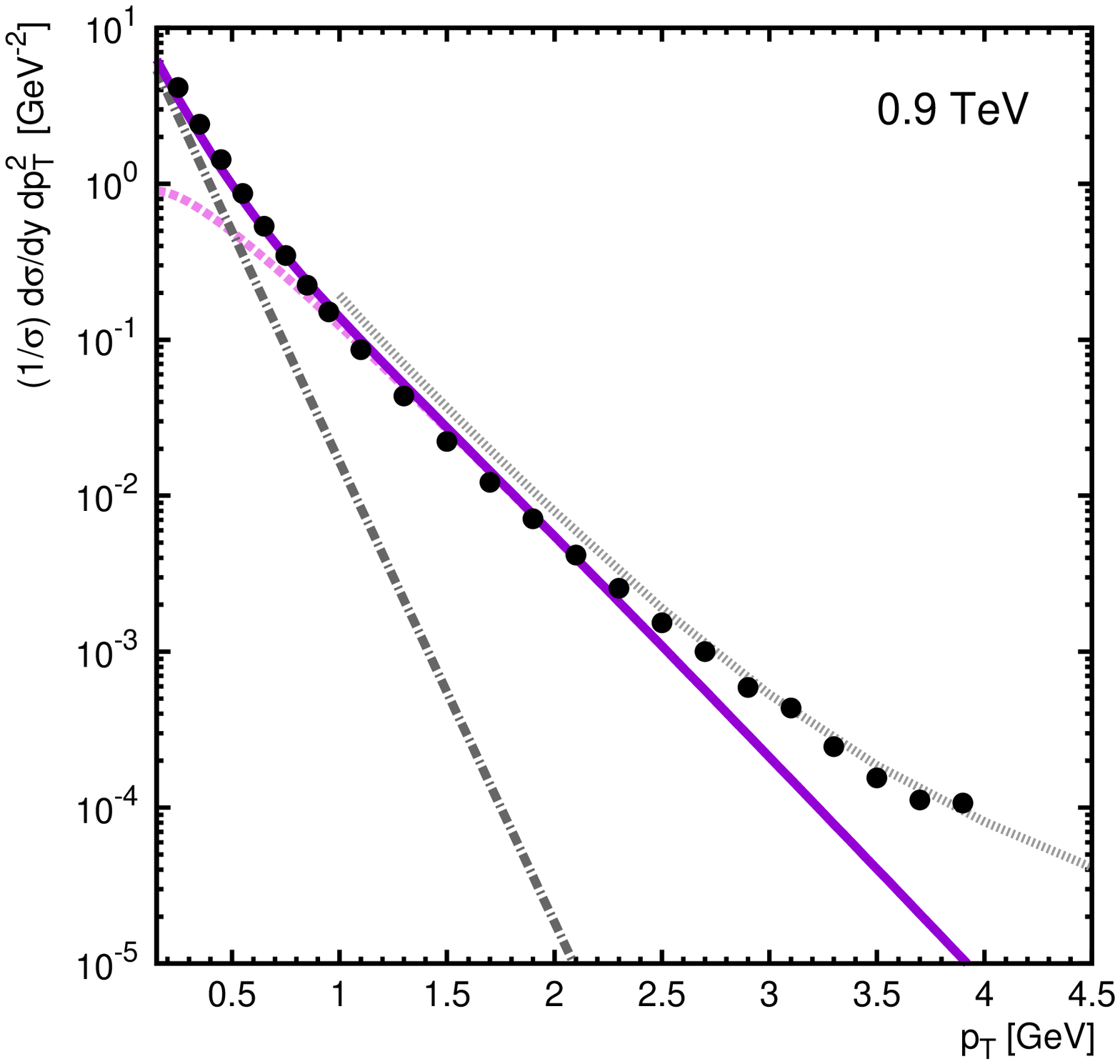, width = 5.25cm, height = 8.0cm, angle = 270}
\vspace{0.7cm} \hspace{-1.1cm}
\epsfig{figure=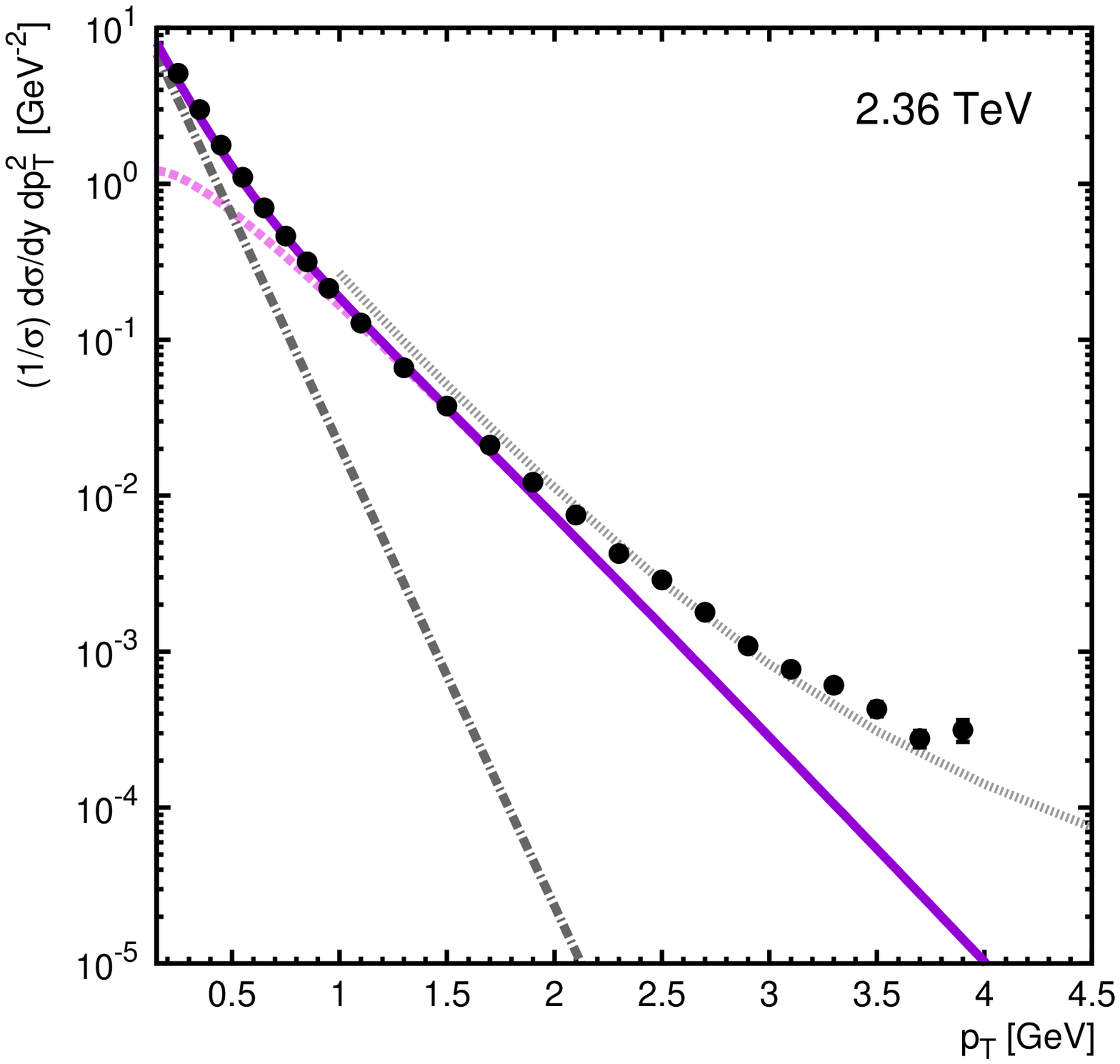, width = 5.25cm, height = 8.0cm, angle = 270}
\epsfig{figure=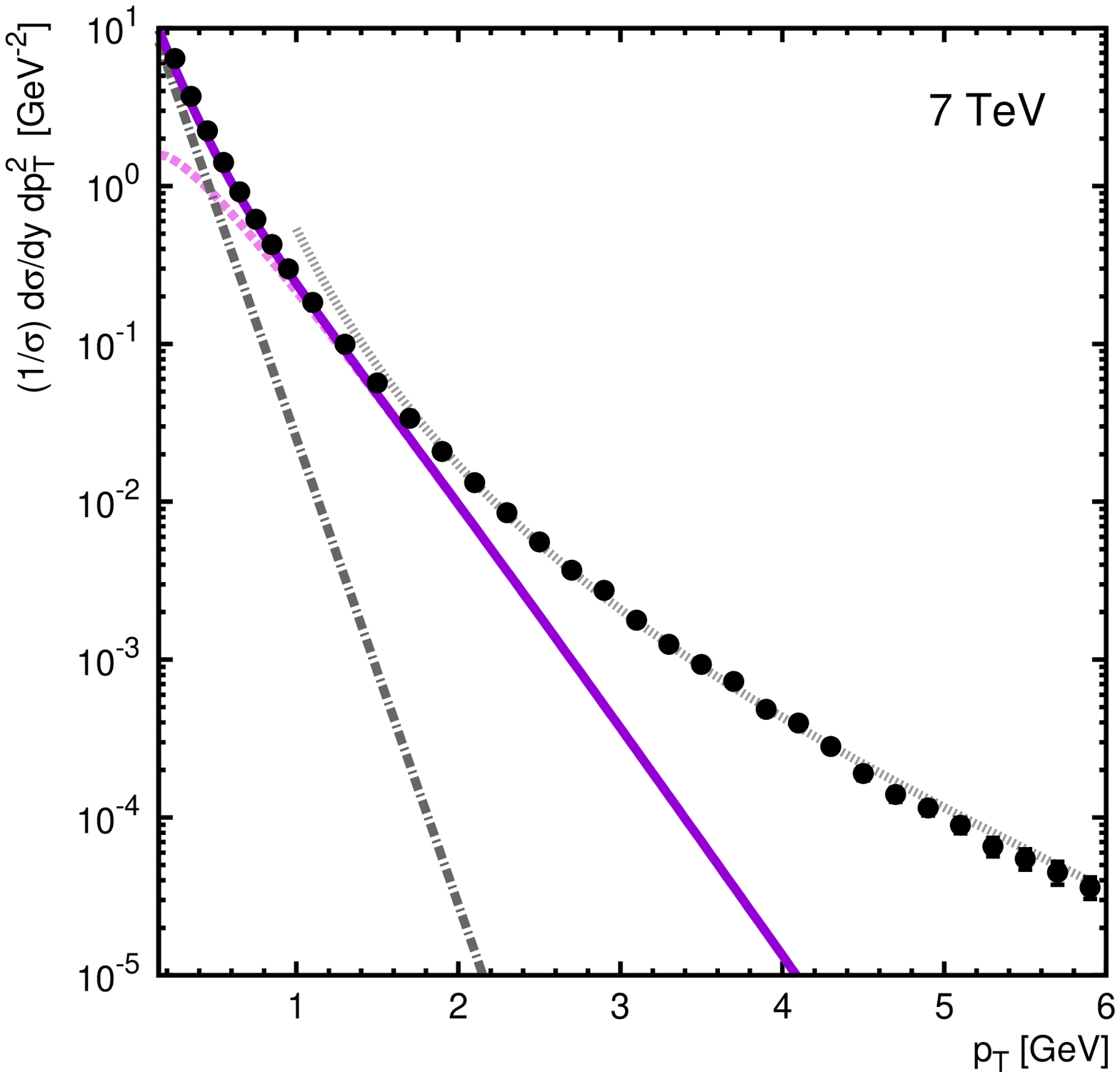, width = 5.25cm, height = 8.0cm, angle = 270}
\caption{The inclusive cross sections of the hadron production in the
$pp$ collisions at the LHC as a function of the transverse momentum.
The dashed and dash-dotted curves correspond to the gluon and quark contributions, respectively.
The solid curves represent their sum. The dotted curves correspond
to the sum of the soft QCD and pQCD predictions as described in the text.
The experimental data are from ATLAS and CMS\cite{23}.}
\label{fig2}
\end{center}
\end{figure}

\begin{figure}
\begin{center}
\epsfig{figure=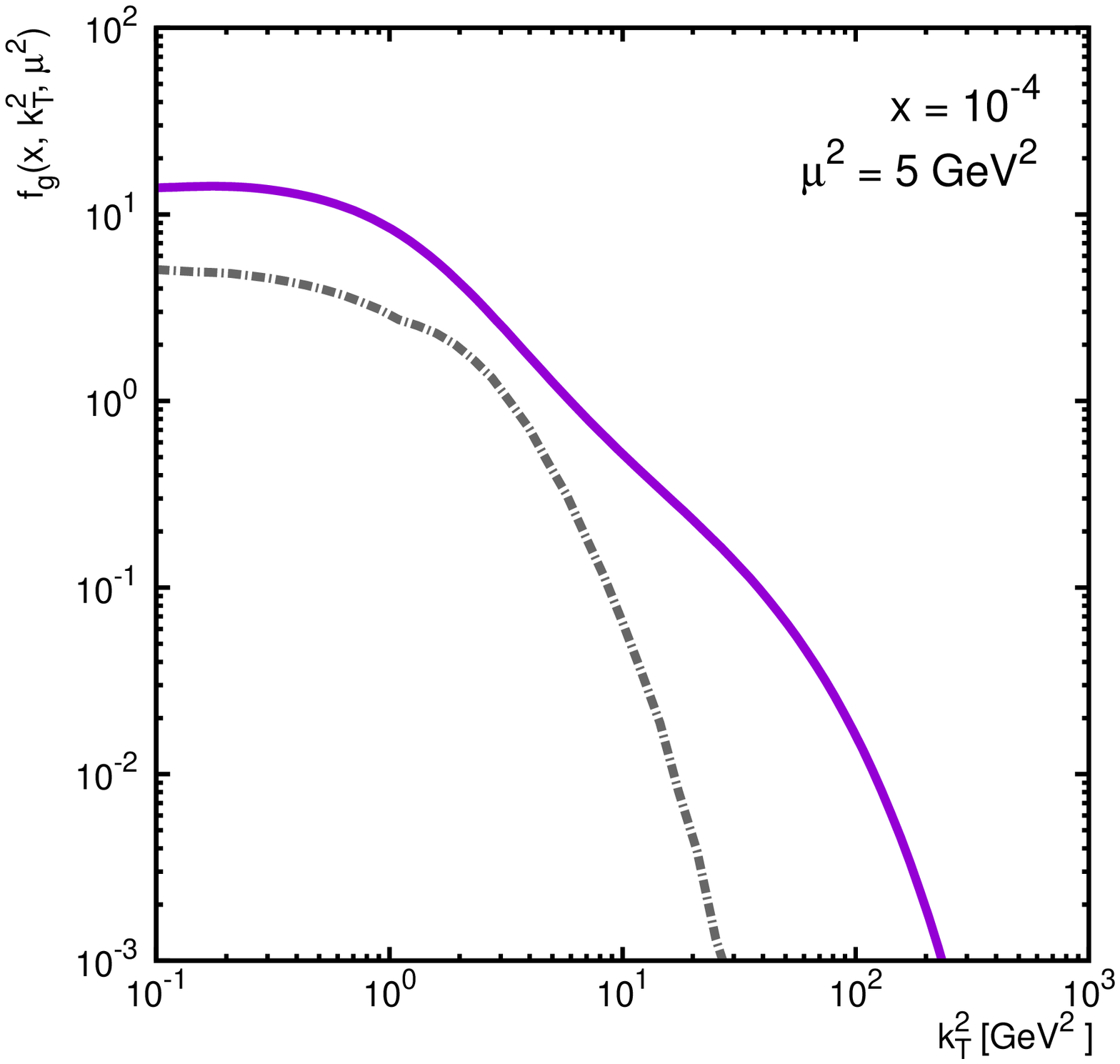, width = 5.25cm, height = 8.0cm, angle = 270}
\vspace{0.7cm} \hspace{-1.1cm}
\epsfig{figure=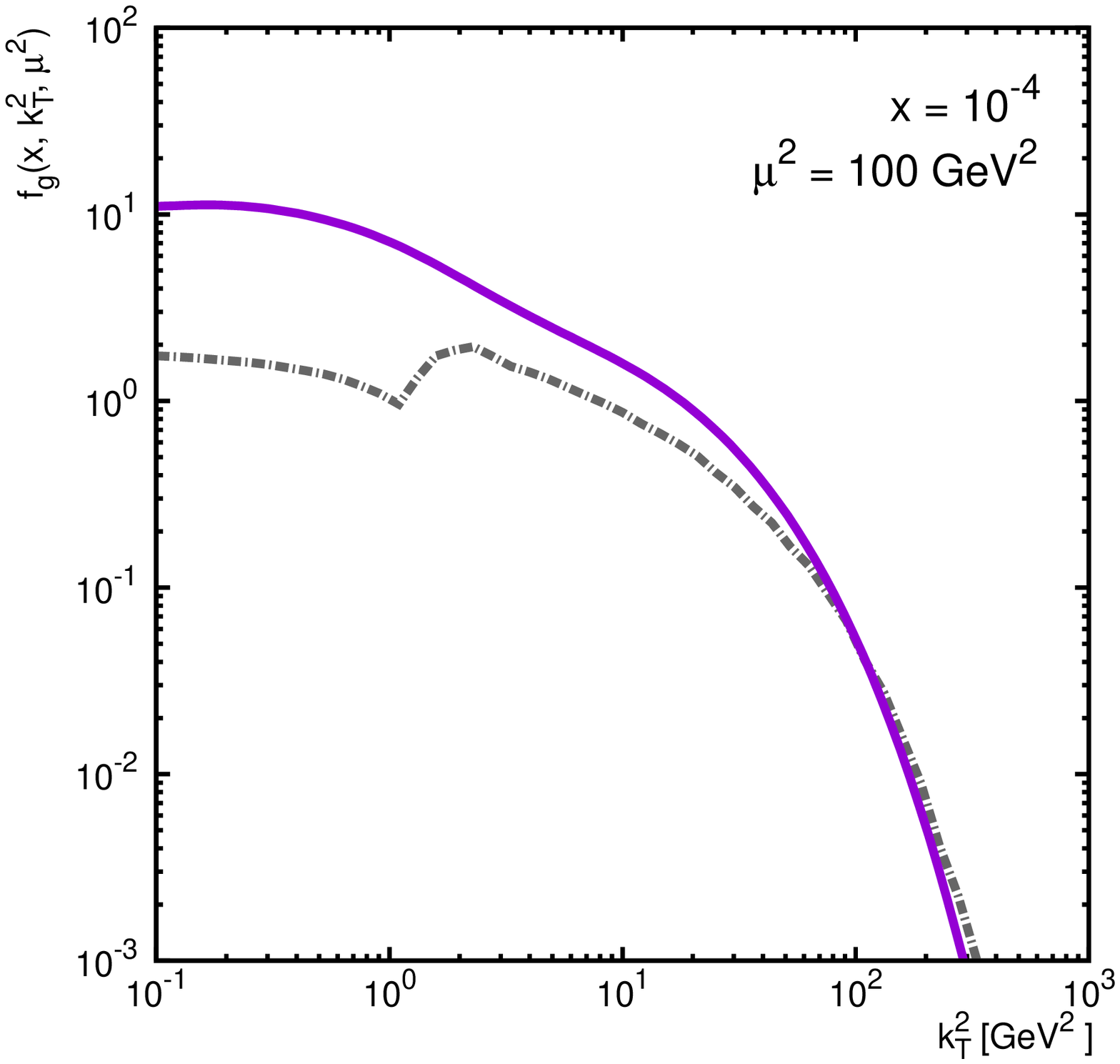, width = 5.25cm, height = 8.0cm, angle = 270} 
\epsfig{figure=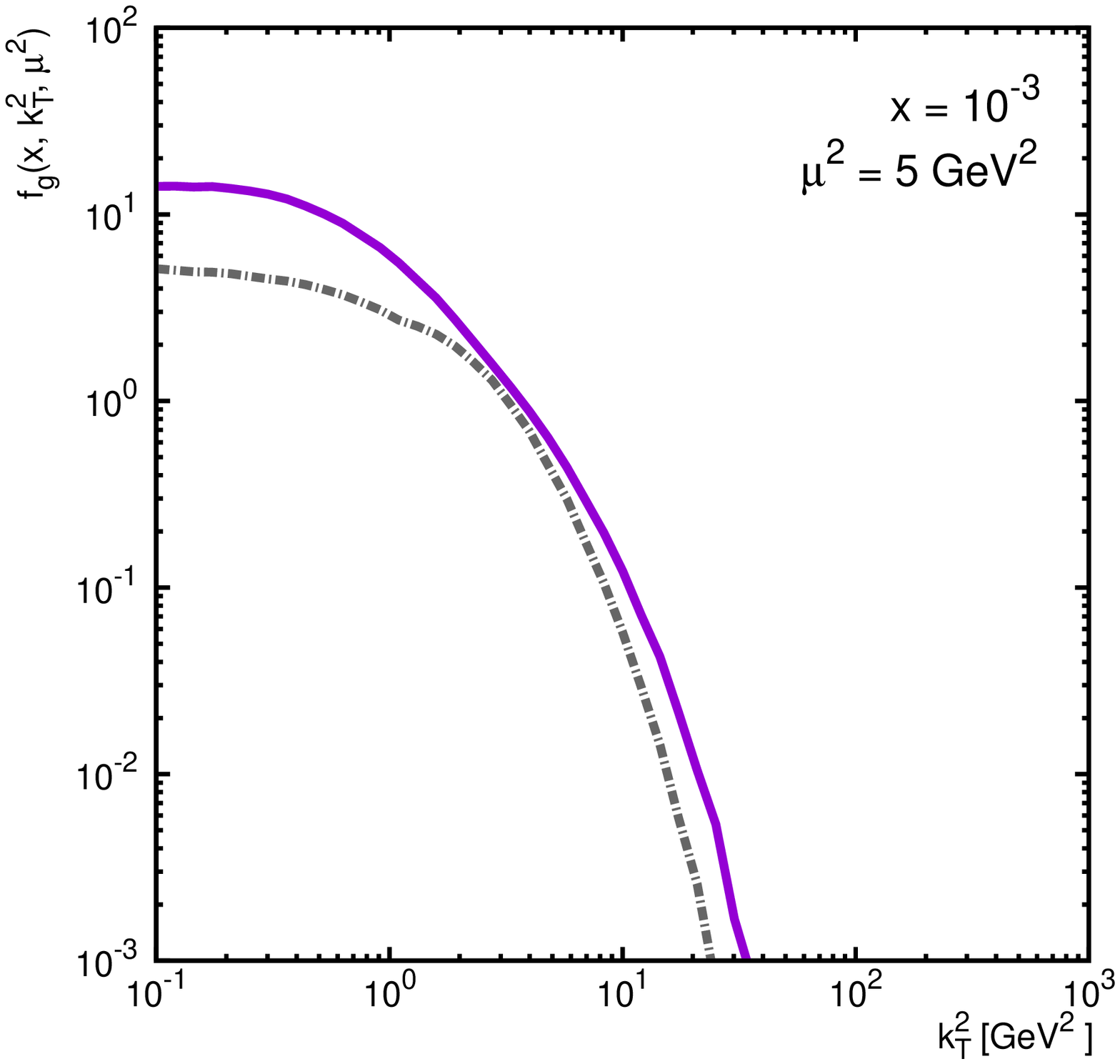, width = 5.25cm, height = 8.0cm, angle = 270}
\hspace{-1.1cm} 
\epsfig{figure=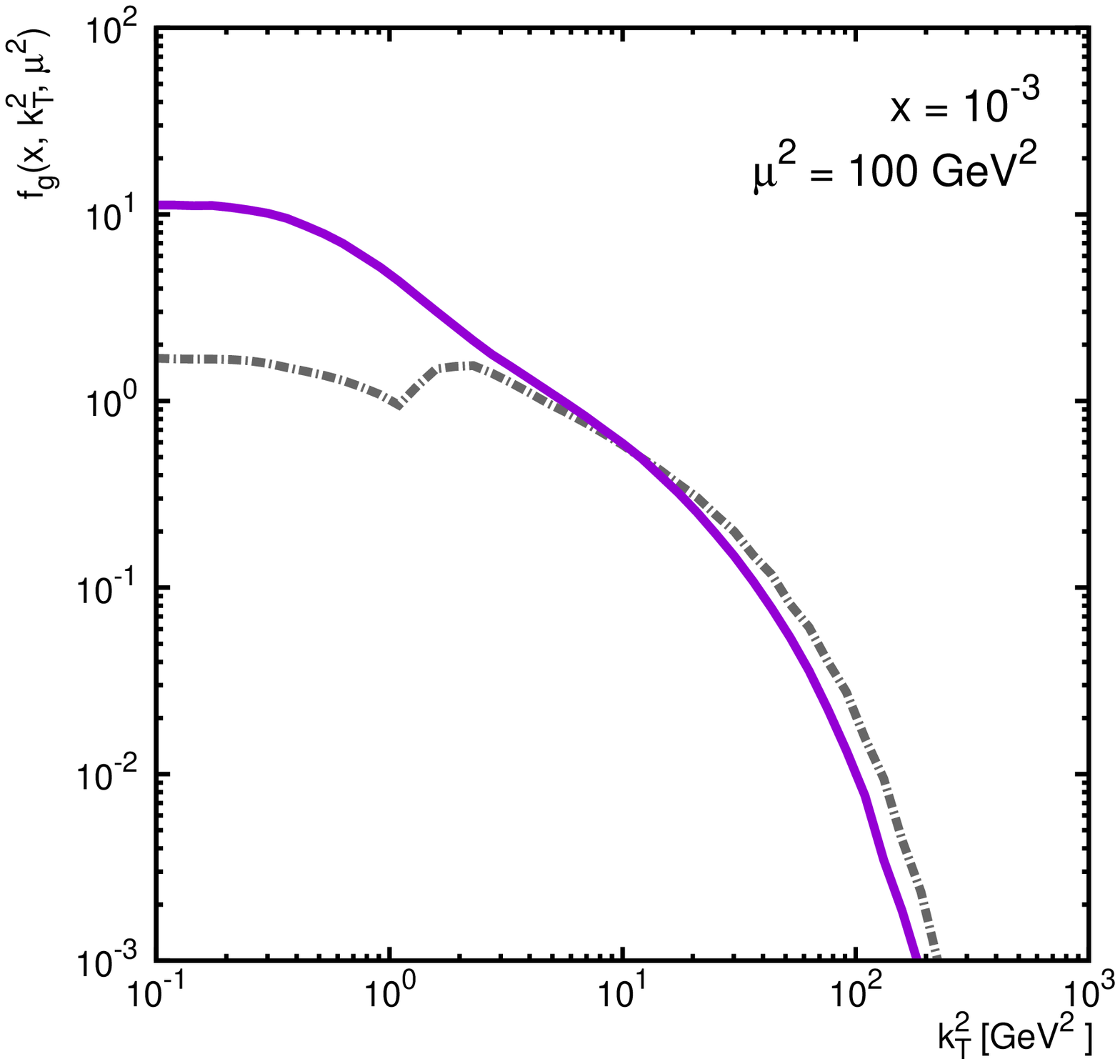, width = 5.25cm, height = 8.0cm, angle = 270}
\caption{The CCFM-evolved TMD gluon densities in the proton calculated 
as a function of the gluon transverse momentum ${\mathbf k_T^2}$ 
at different $x$ and $\mu^2$. The solid and dashed curves
correspond to the proposed MD'2015 and A0 gluon densities, respectively.}
\label{fig3}
\end{center}
\end{figure}

\begin{figure}
\begin{center}
\epsfig{figure=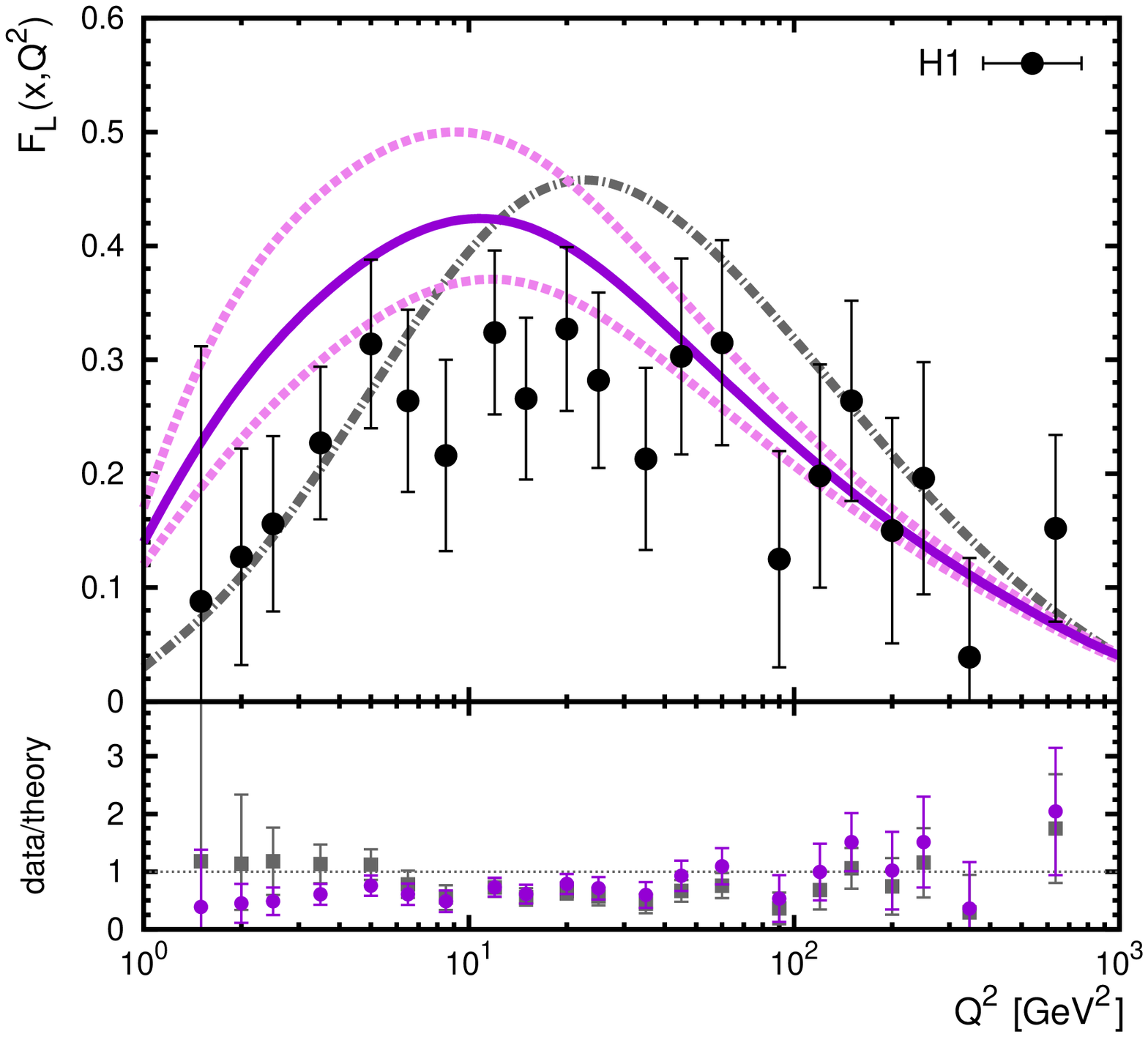, width = 5.5cm, angle = 270}
\vspace{0.7cm} \hspace{-1cm}
\epsfig{figure=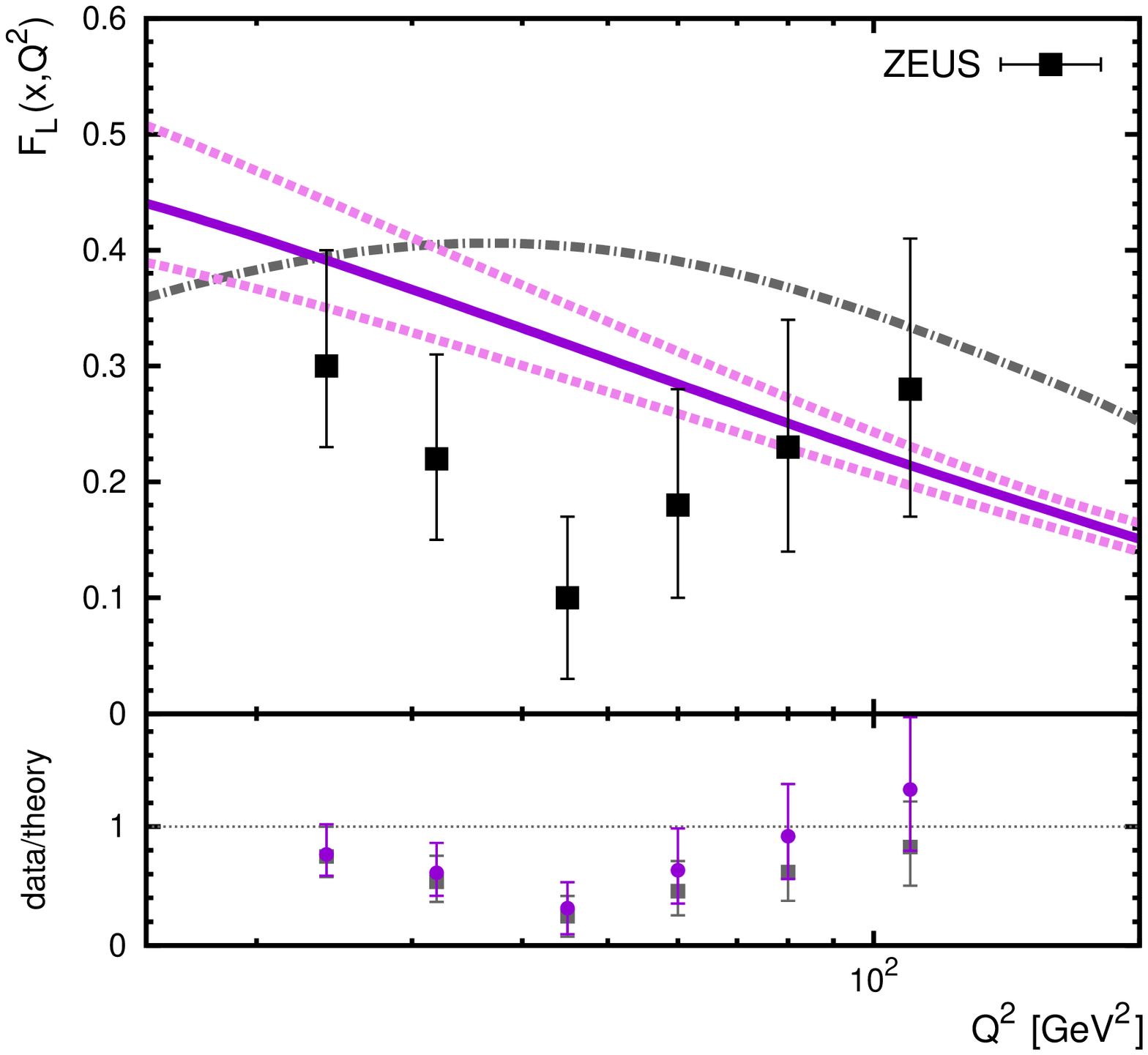, width = 5.5cm, angle = 270}
\caption{The longitudinal proton structure function $F_L(x,Q^2)$ 
as a function of $Q^2$. The solid curve corresponds to the predictions obtained with the 
proposed MD'2015 gluon density. The upper and lower dashed curves correspond to the usual 
scale variations in these calculations. The dash-dotted curve represents the results
obtained with the A0 gluon. The experimental data are from H1\cite{41} and ZEUS\cite{42}.
In the ZEUS measurements the ratio $Q^2/x$ is a constant for each bin, which corresponds to
$y = 0.71$ and $\sqrt s = 225$~GeV, where $y = Q^2/xs$.}
\label{fig4}
\end{center}
\end{figure}

\begin{figure}
\begin{center}
\epsfig{figure=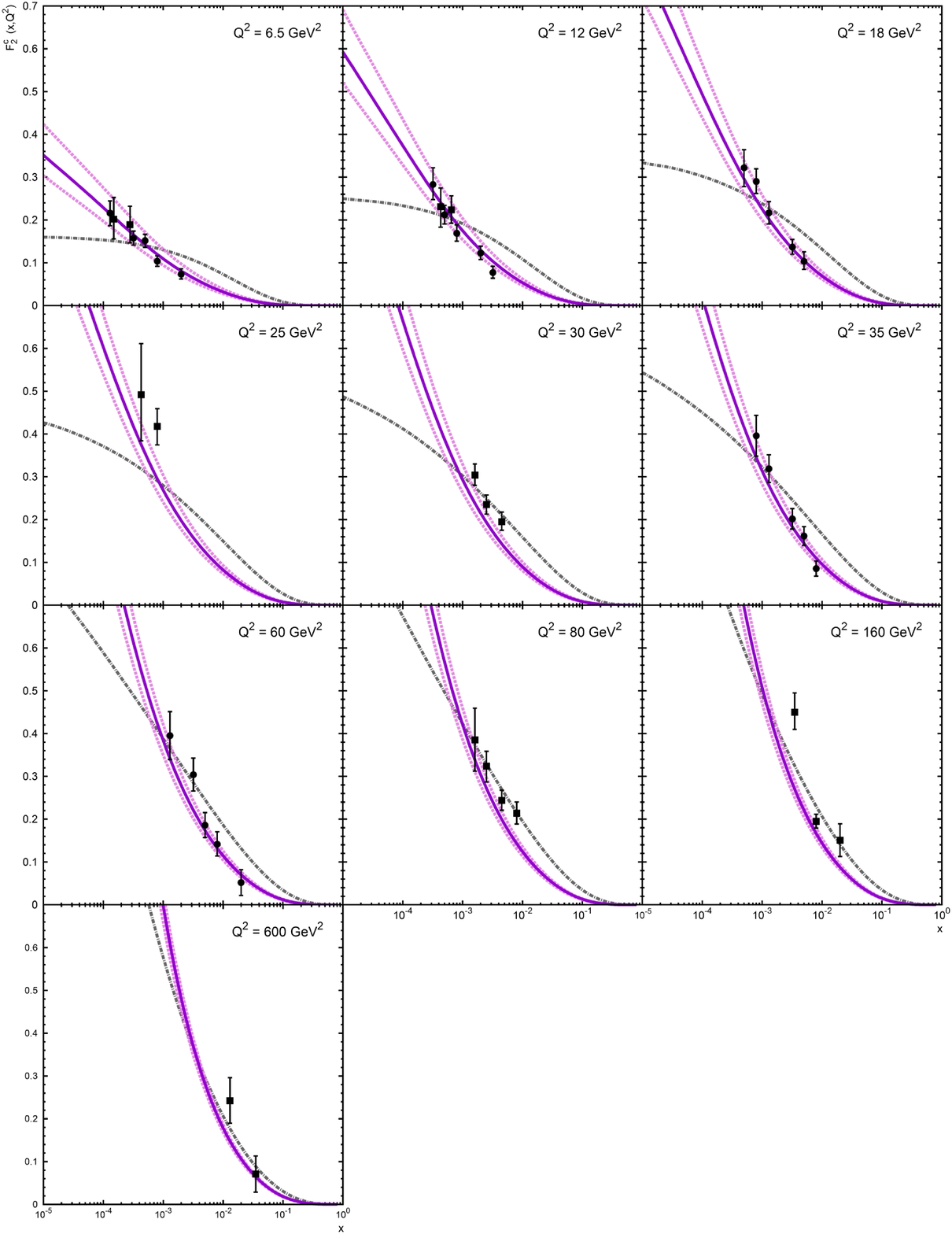, width = 15cm}
\caption{The charm contribution to the structure function $F_2(x,Q^2)$ 
as a function of $x$ calculated at different $Q^2$. Notation of all curves is the same as in Fig.~4. 
The experimental data are from ZEUS\cite{43} and H1\cite{44}.}
\label{fig5}
\end{center}
\end{figure}

\begin{figure}
\begin{center}
\epsfig{figure=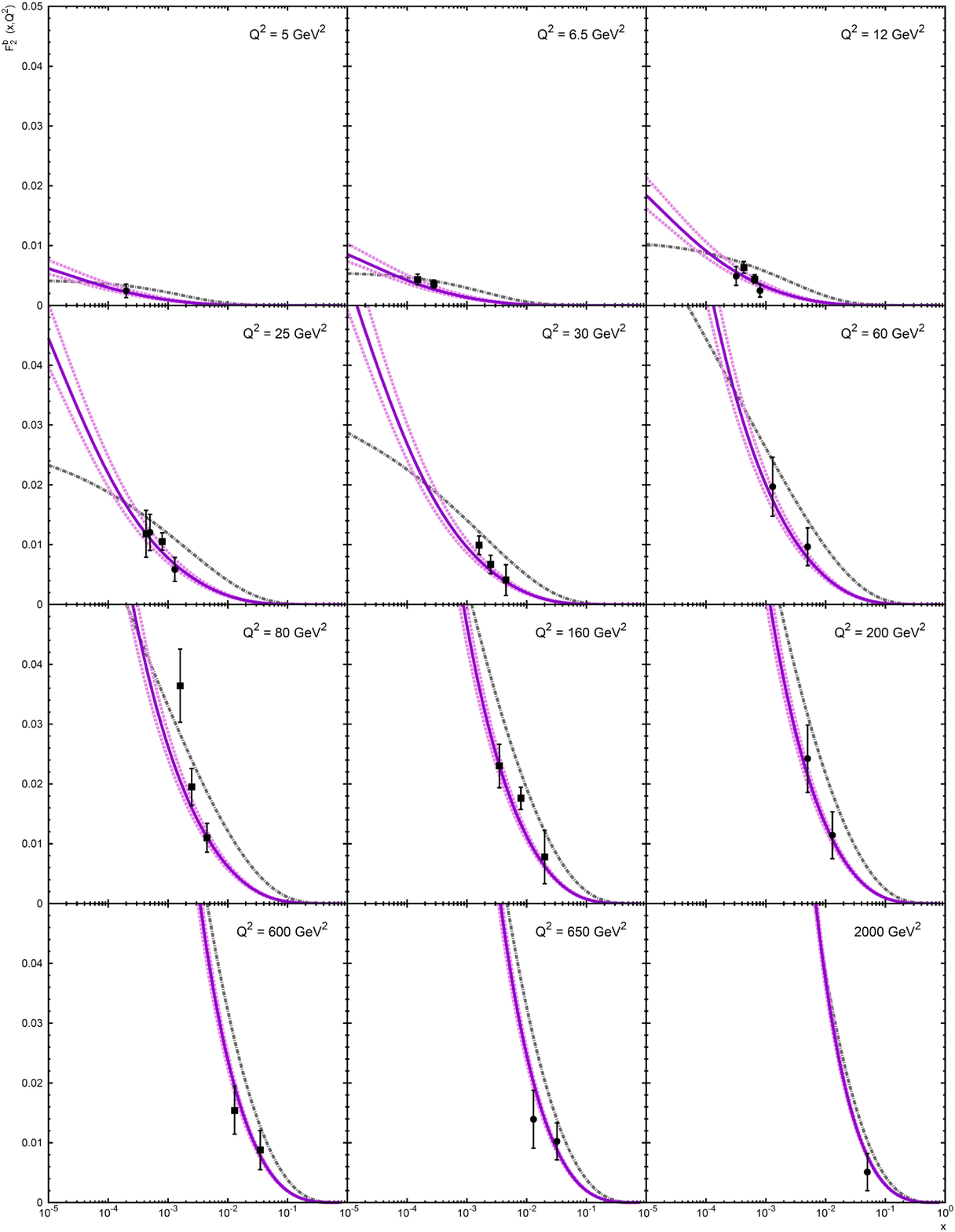, width = 15cm}
\caption{The beauty contribution to the structure function $F_2(x,Q^2)$ 
as a function of $x$ calculated at different $Q^2$. Notation of all curves is the same as in Fig.~4. 
The experimental data are from ZEUS\cite{43} and H1\cite{45}.}
\label{fig6}
\end{center}
\end{figure}

\begin{figure}
\begin{center}
\epsfig{figure=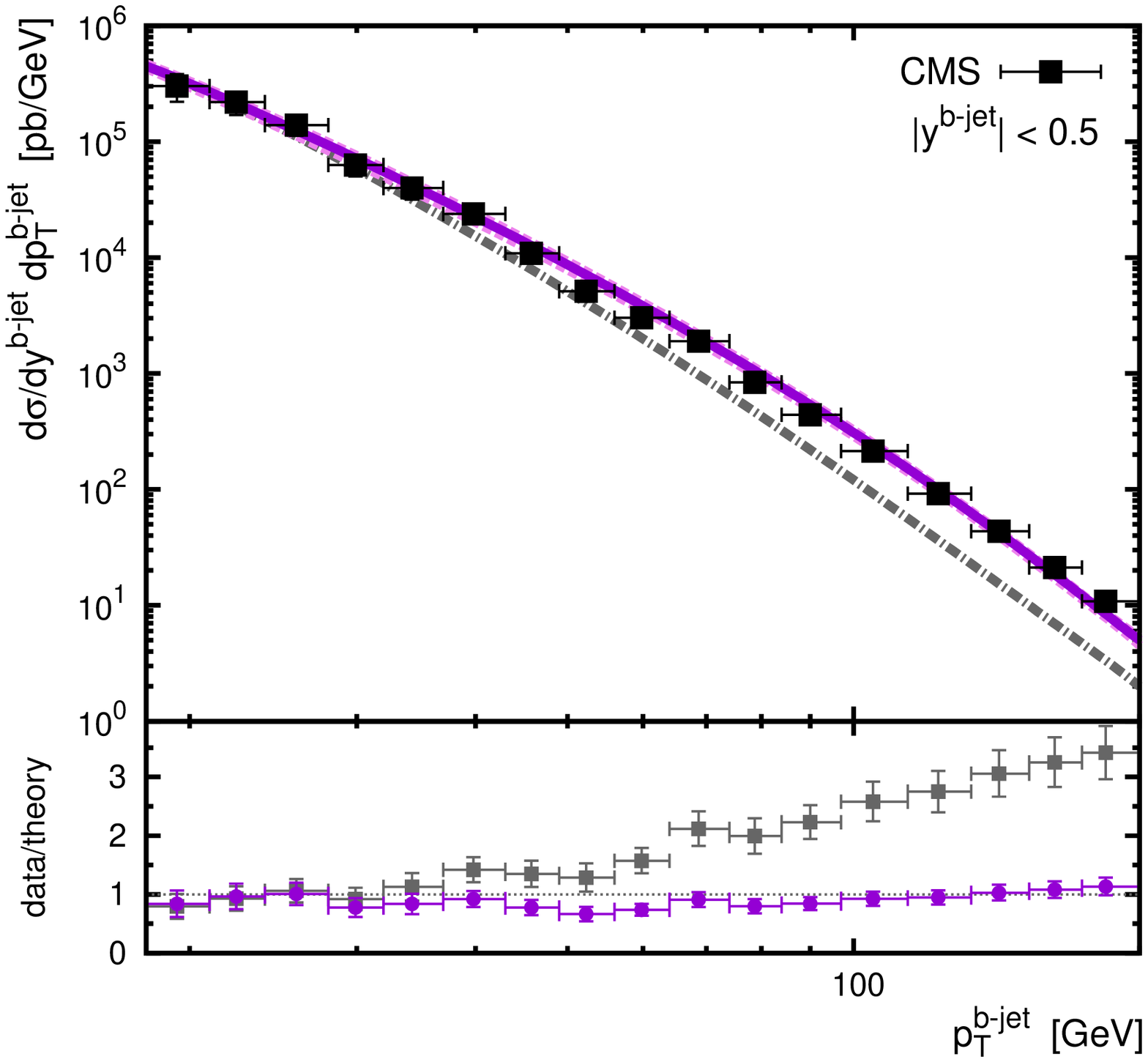, width = 5.5cm, angle = 270}
\vspace{0.7cm} \hspace{-1cm}
\epsfig{figure=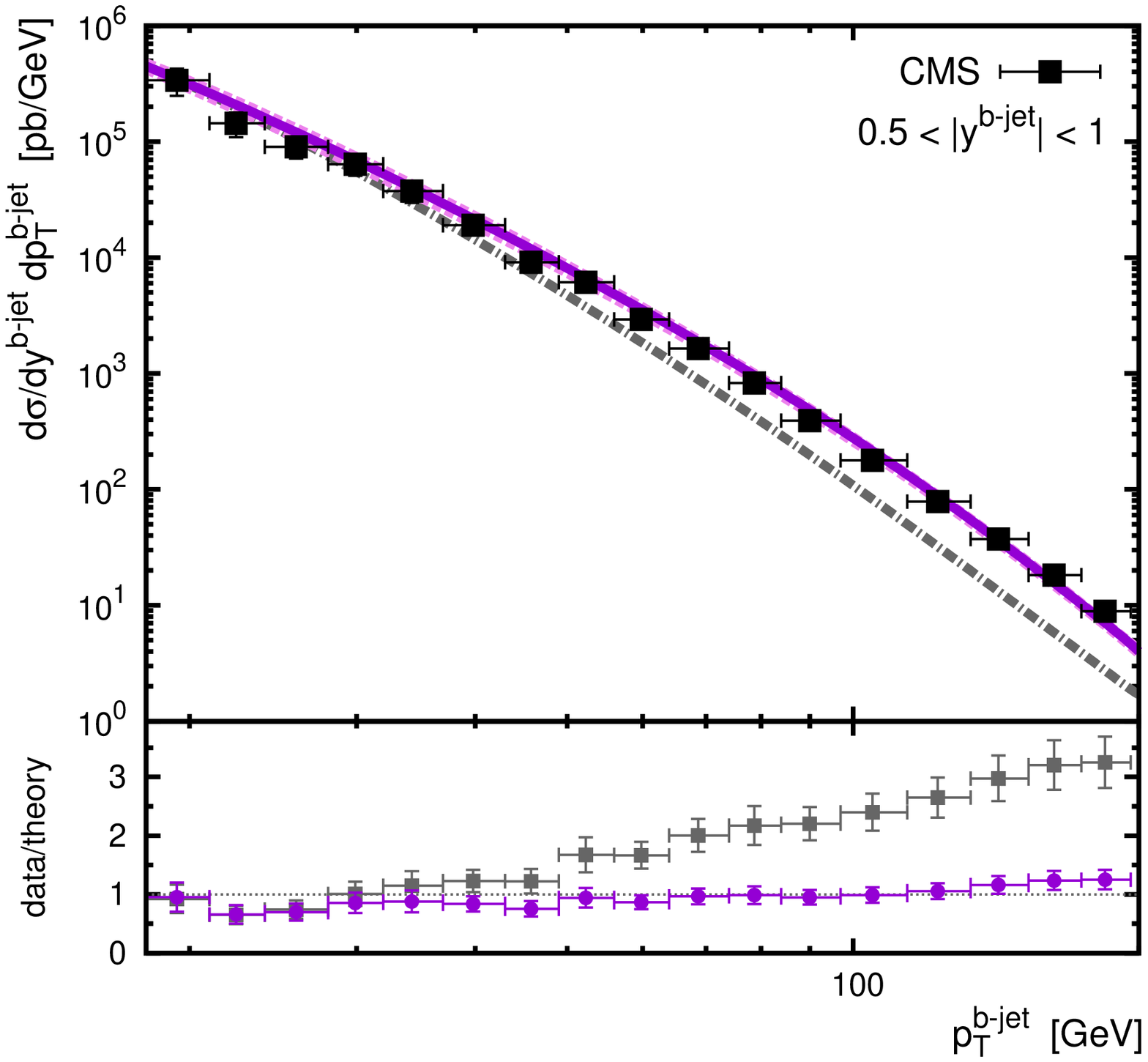, width = 5.5cm, angle = 270}
\vspace{0.7cm}
\epsfig{figure=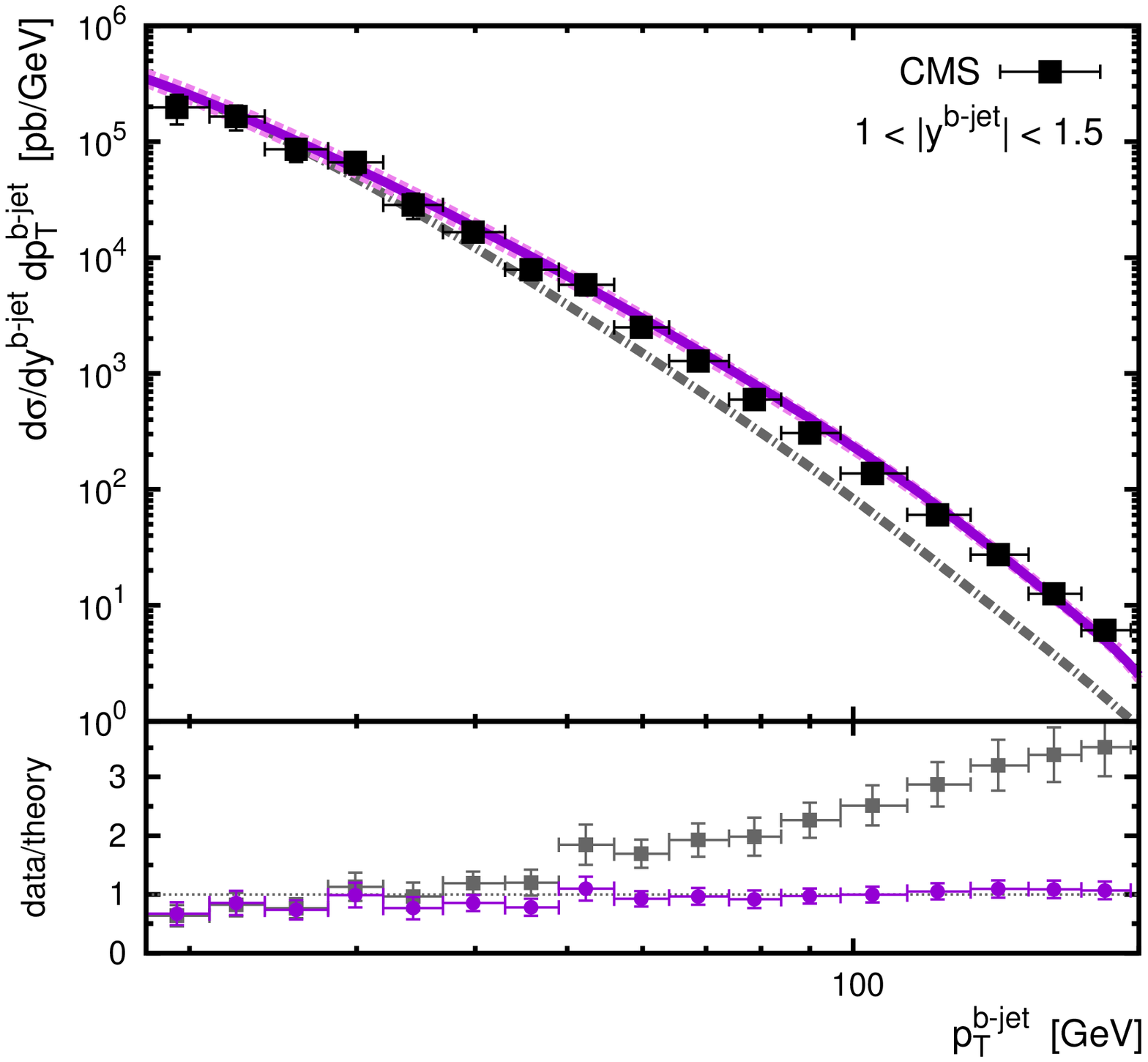, width = 5.5cm, angle = 270}
\hspace{-1cm}
\epsfig{figure=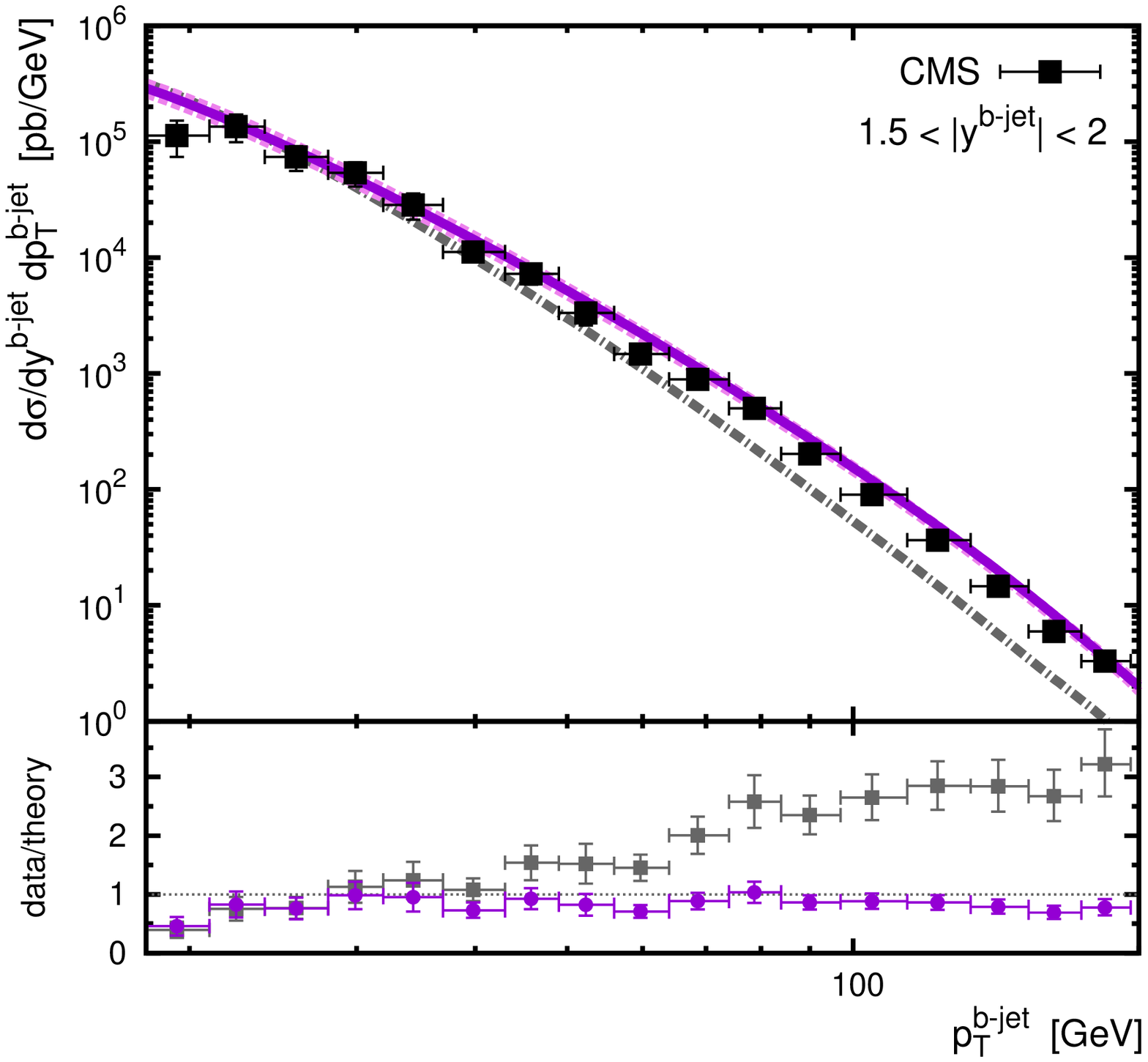, width = 5.5cm, angle = 270}
\epsfig{figure=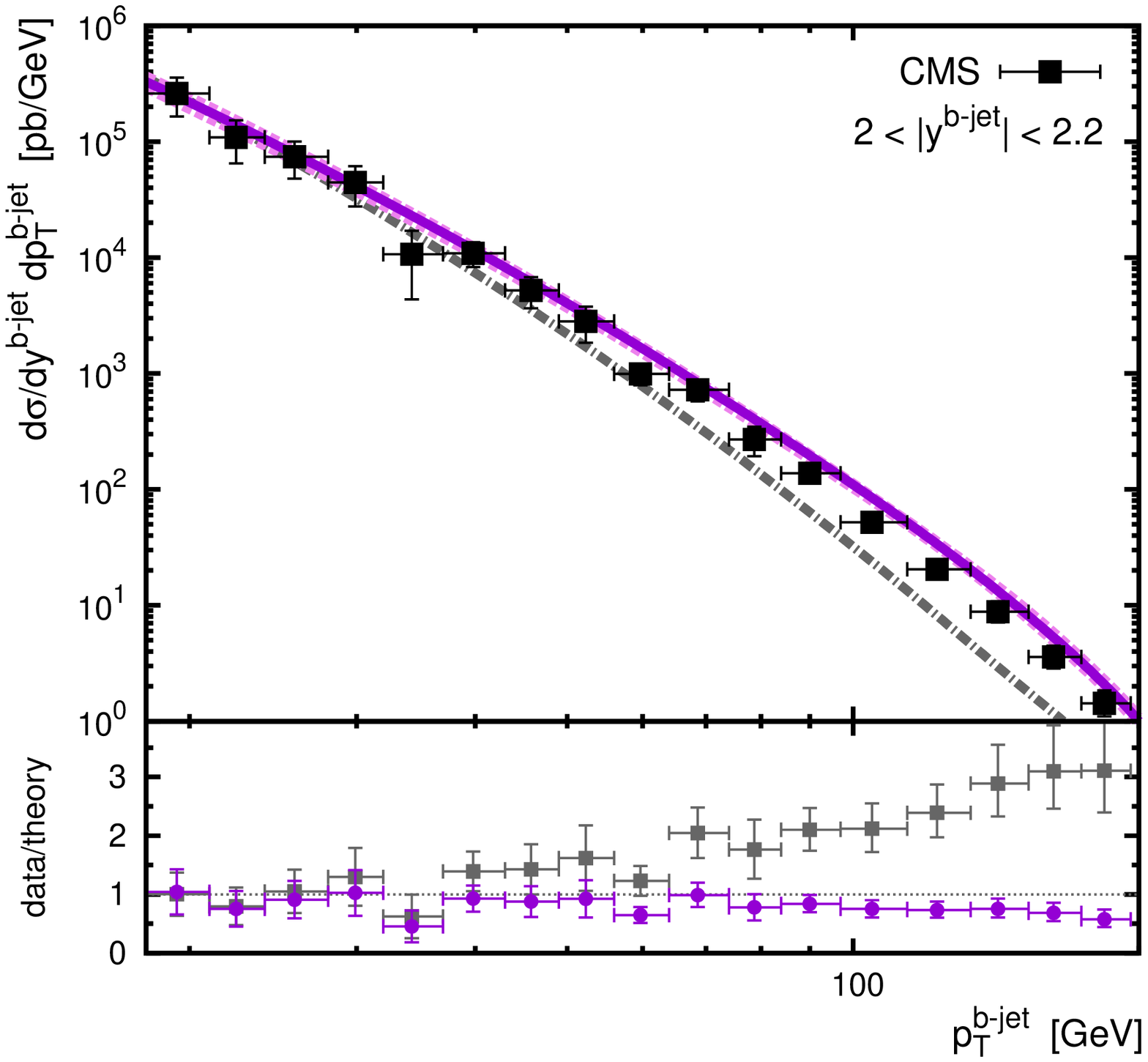, width = 5.5cm, angle = 270}
\caption{The double differential cross sections $d\sigma/dy dp_T$
of the inclusive $b$-jet production as a function of the leading 
jet transverse momentum in different $y$ regions. 
The used kinematical cuts are described in the text. 
Notation of all curves is the same as in Fig.~4. 
The experimental data are from CMS\cite{46}.}
\label{fig7}
\end{center}
\end{figure}

\begin{figure}
\begin{center}
\epsfig{figure=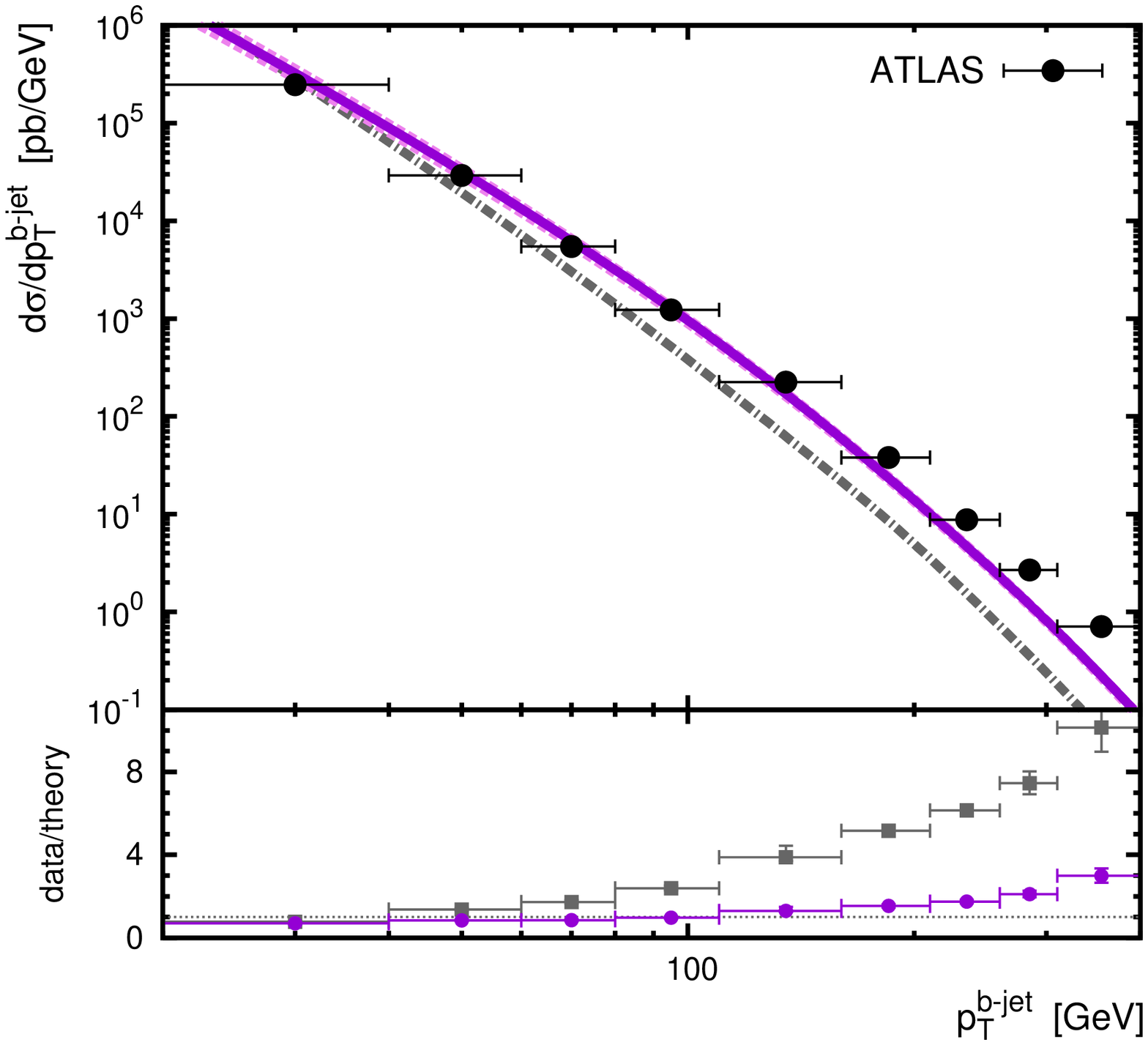, width = 5.5cm, angle = 270}
\vspace{0.7cm} \hspace{-1cm}
\epsfig{figure=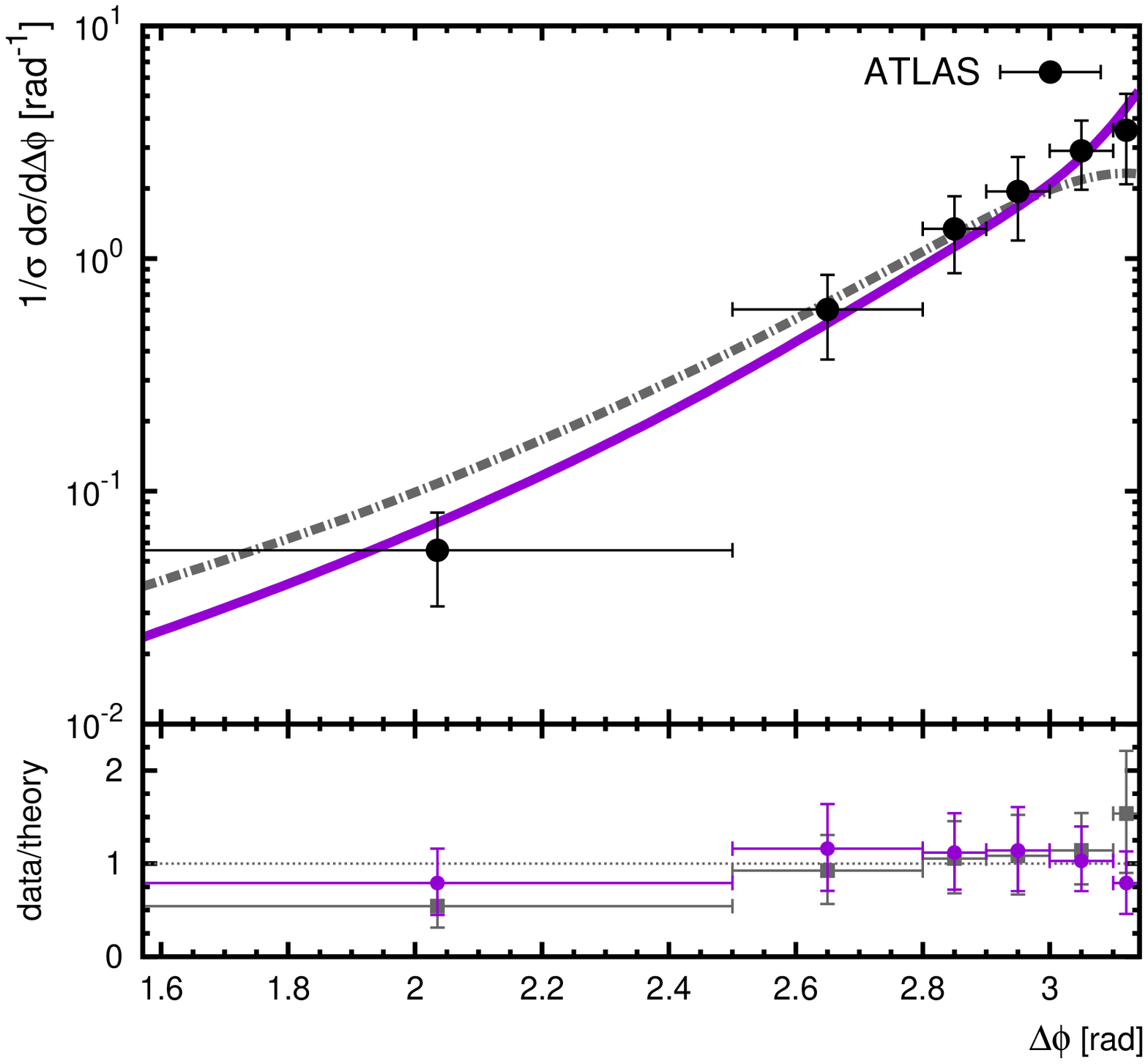, width = 5.5cm, angle = 270}
\caption{The distributions in the leading $b$-jet transverse momentum and 
azimuthal angle difference between the $b$-jets produced at the LHC.
The used kinematical cuts are described in the text. 
Notation of all curves is the same as in Fig.~4. 
The experimental data are from ATLAS\cite{47}.}
\label{fig8}
\end{center}
\end{figure}

\begin{figure}
\begin{center}
\epsfig{figure=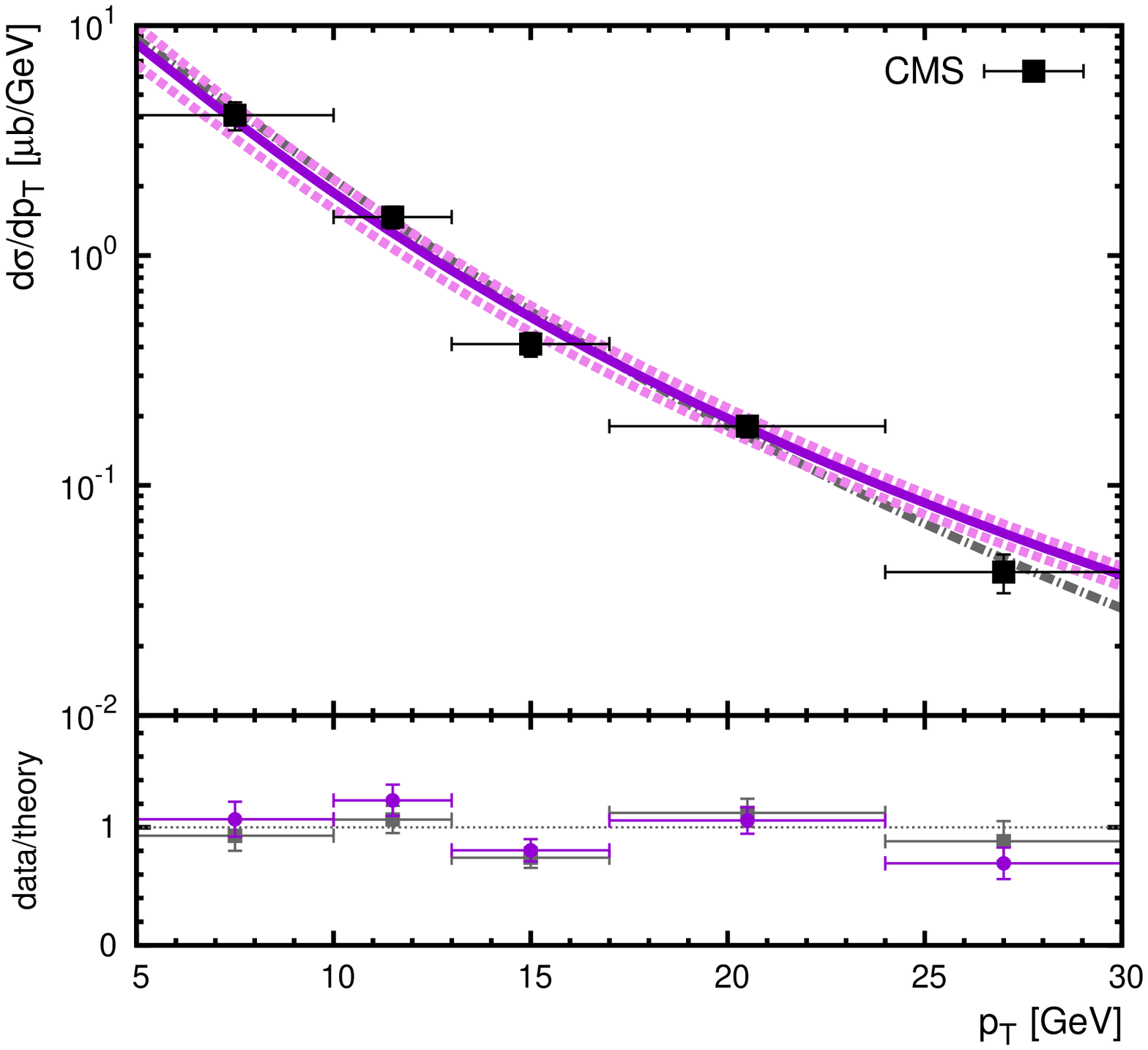, width = 5.5cm, angle = 270}
\vspace{0.7cm} \hspace{-1cm}
\epsfig{figure=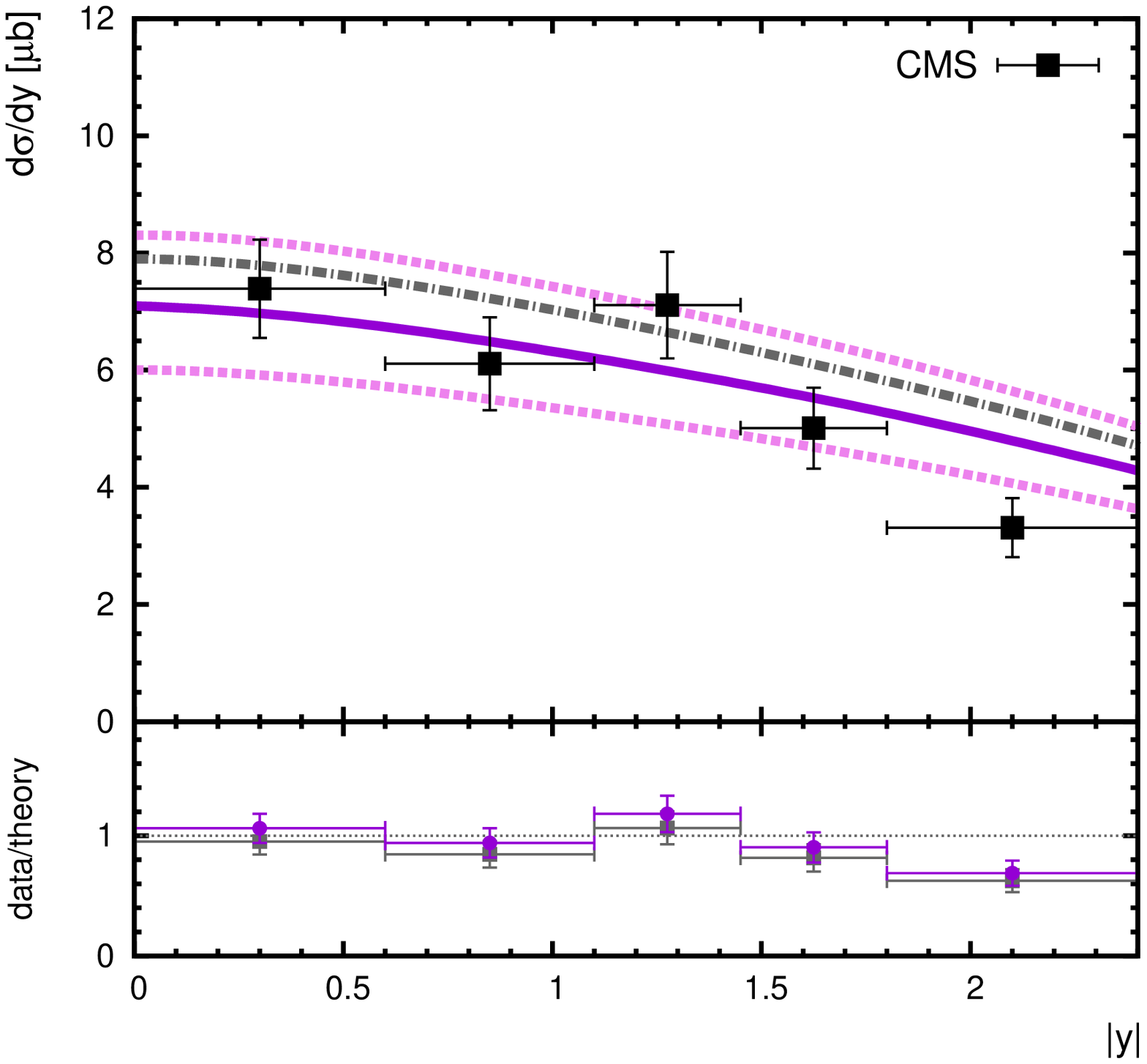, width = 5.5cm, angle = 270}
\caption{The transverse momentum and rapidity distributions of the $B^+$ meson production at
the LHC. The used kinematical cuts are described in the text. 
Notation of all curves is the same as in Fig.~4. 
The experimental data are from CMS\cite{48}.}
\label{fig9}
\end{center}
\end{figure}

\begin{figure}
\begin{center}
\epsfig{figure=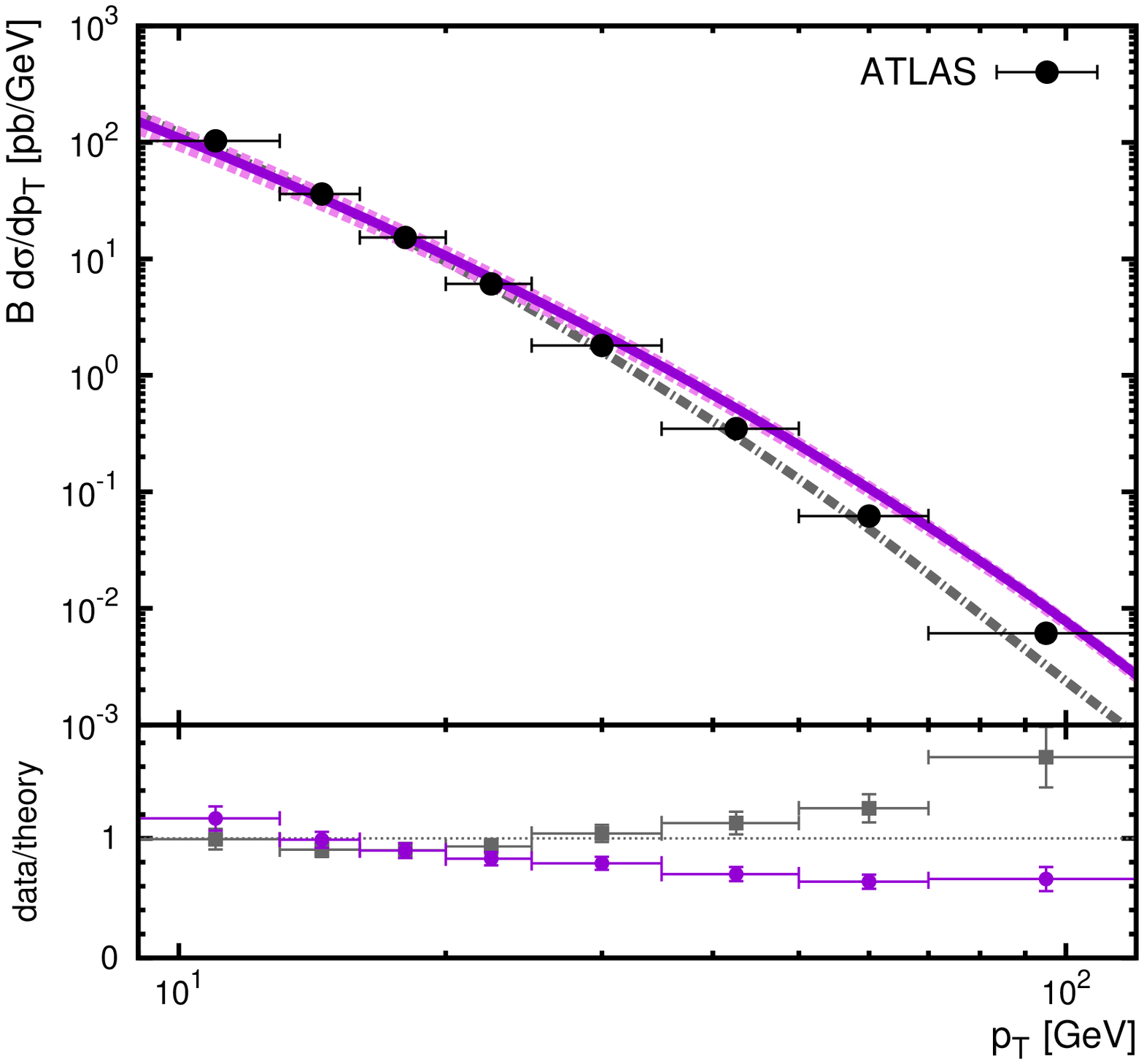, width = 5.5cm, angle = 270}
\vspace{0.7cm} \hspace{-1cm}
\epsfig{figure=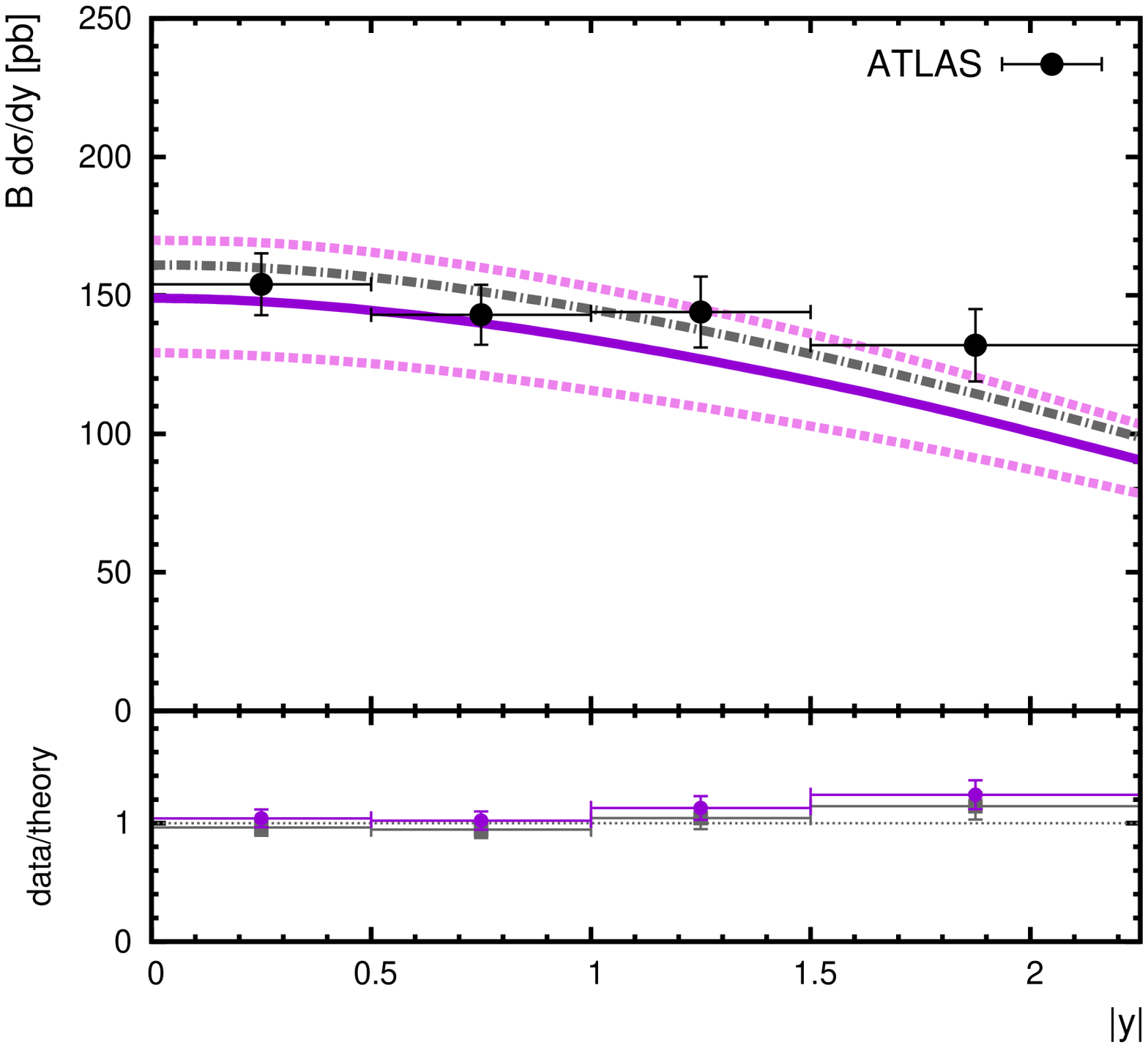, width = 5.5cm, angle = 270}
\caption{The transverse momentum and rapidity distributions of the $B^+$ meson production at
the LHC. The used kinematical cuts are described in the text. 
Notation of all curves is the same as in Fig.~4. 
The experimental data are from ATLAS\cite{49}.}
\label{fig10}
\end{center}
\end{figure}

\begin{figure}
\begin{center}
\epsfig{figure=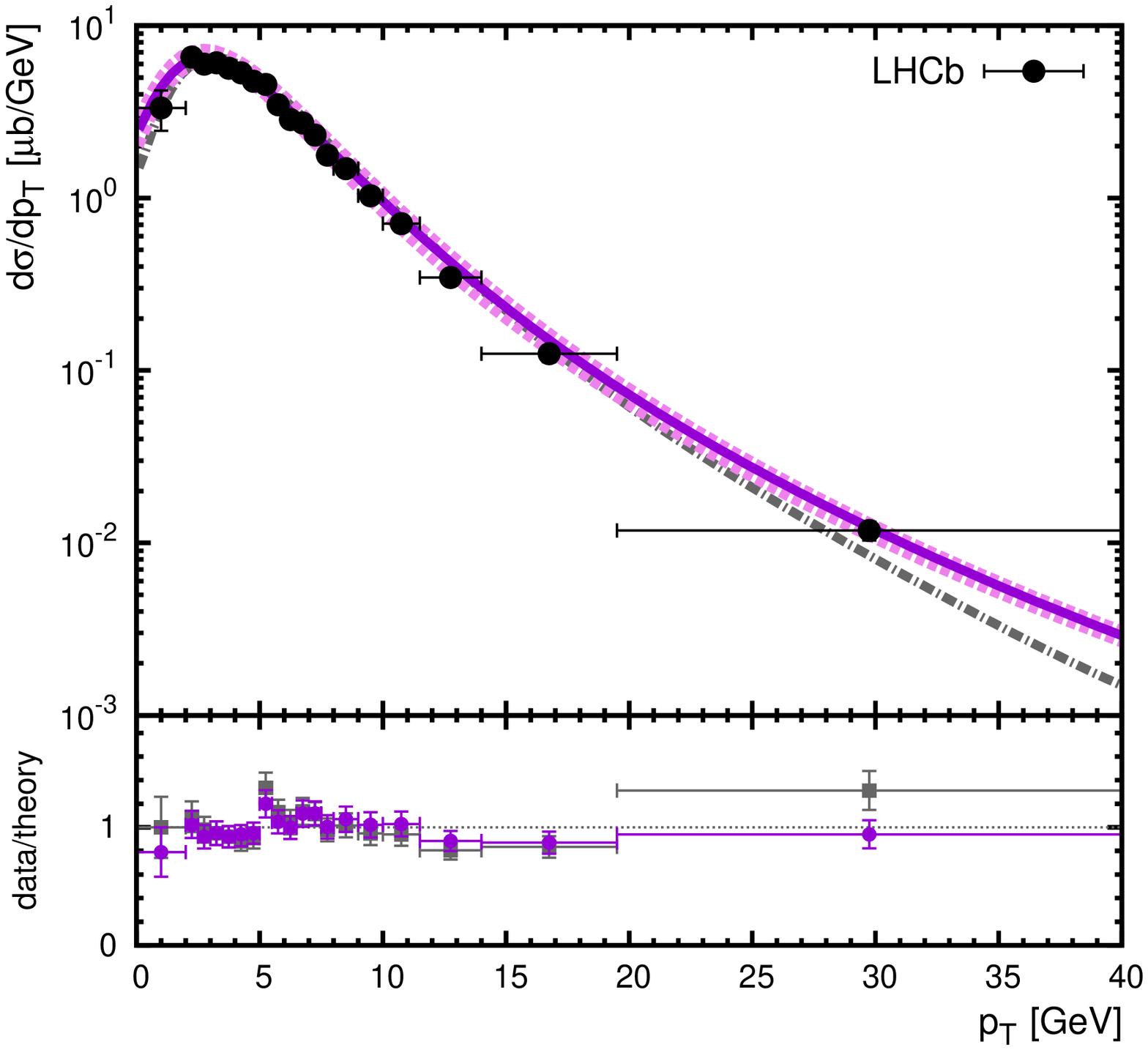, width = 5.5cm, angle = 270}
\vspace{0.7cm} \hspace{-1cm}
\epsfig{figure=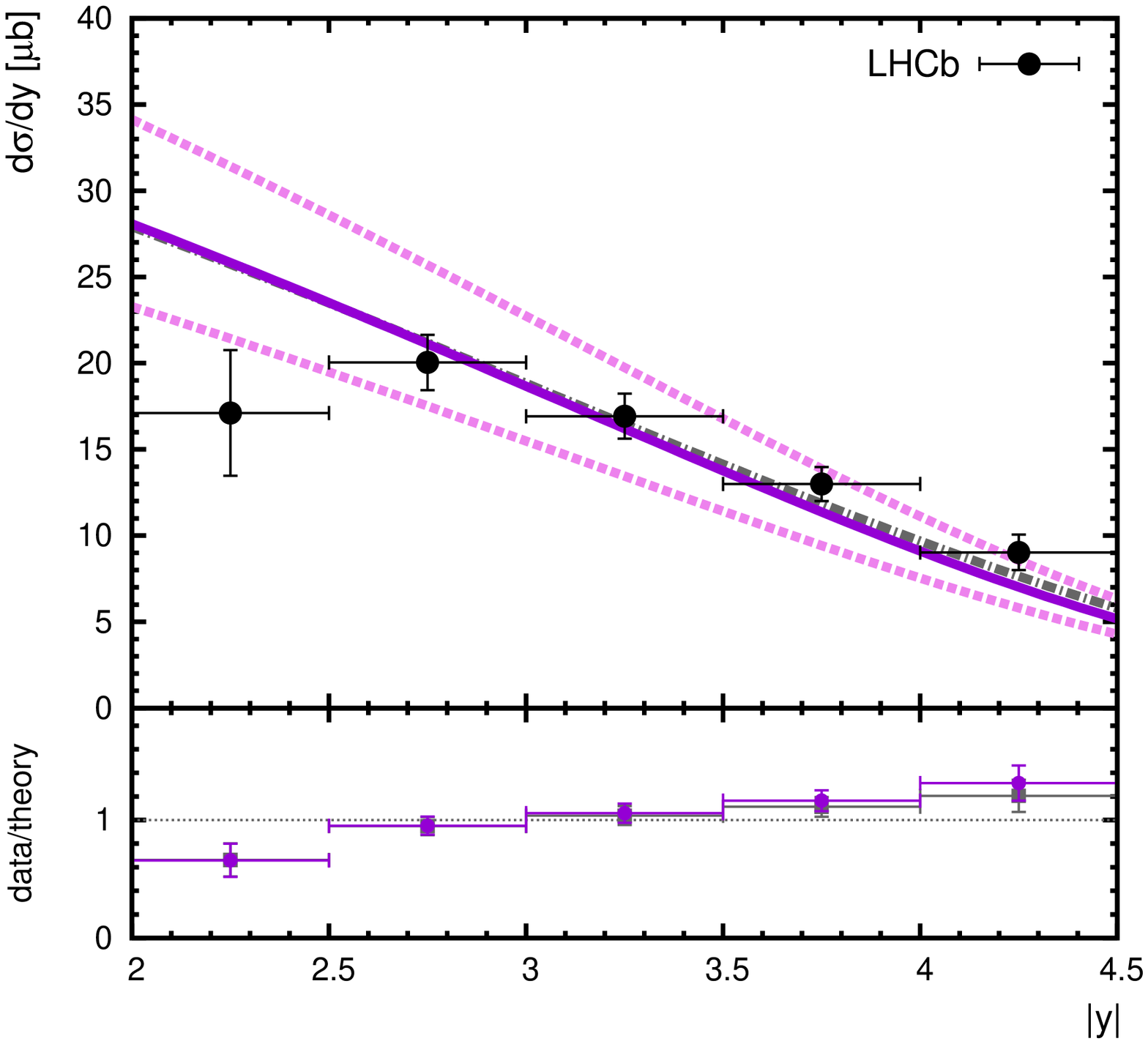, width = 5.5cm, angle = 270}
\caption{The transverse momentum and rapidity distributions of the $B^\pm$ meson production at
the LHC. The used kinematical cuts are described in the text. 
Notation of all curves is the same as in Fig.~4. 
The experimental data are from LHCb\cite{50}.}
\label{fig11}
\end{center}
\end{figure}

\begin{figure}
\begin{center}
\epsfig{figure=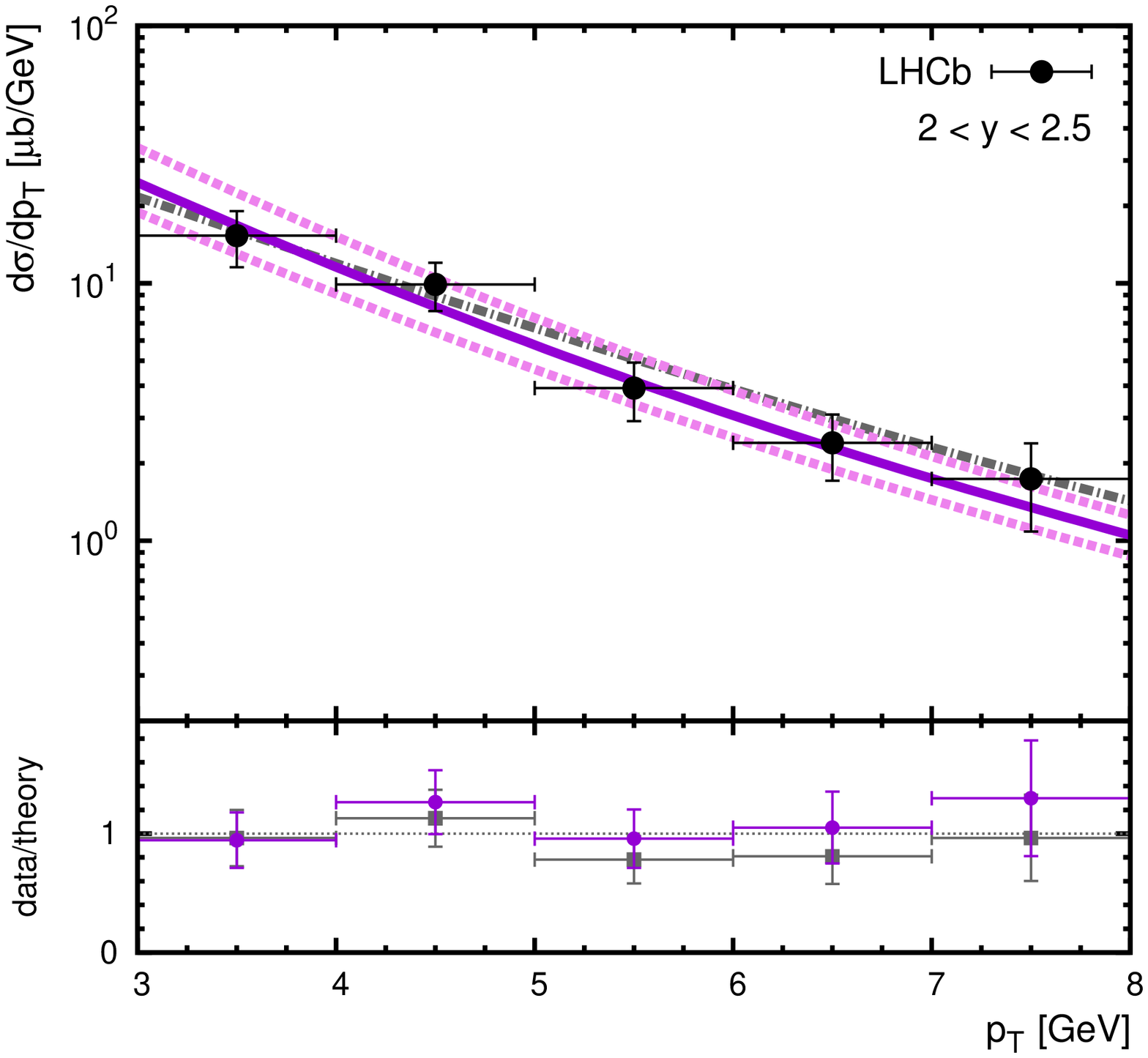, width = 5.5cm, angle = 270}
\vspace{0.7cm} \hspace{-1cm}
\epsfig{figure=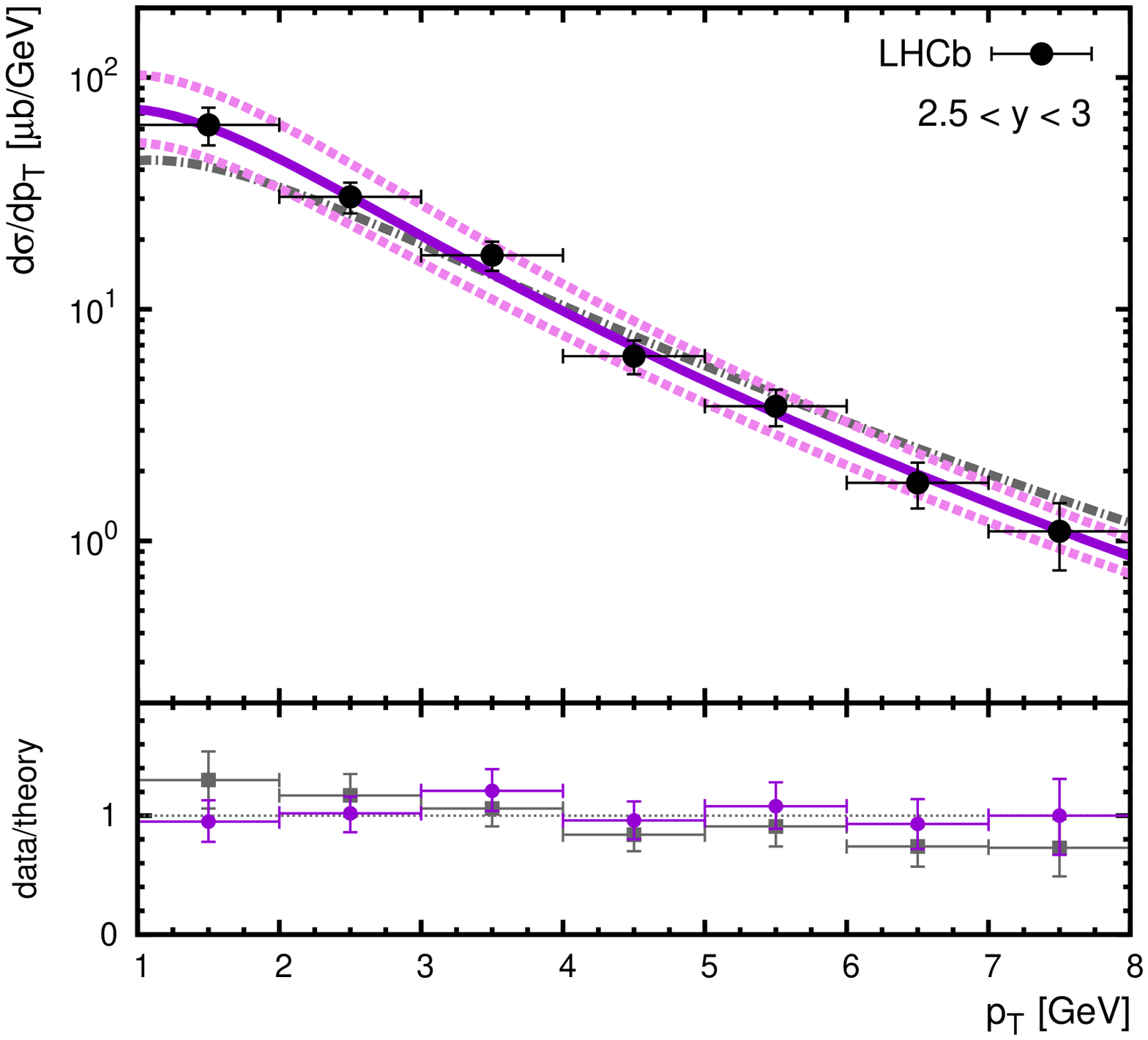, width = 5.5cm, angle = 270}
\vspace{0.7cm}
\epsfig{figure=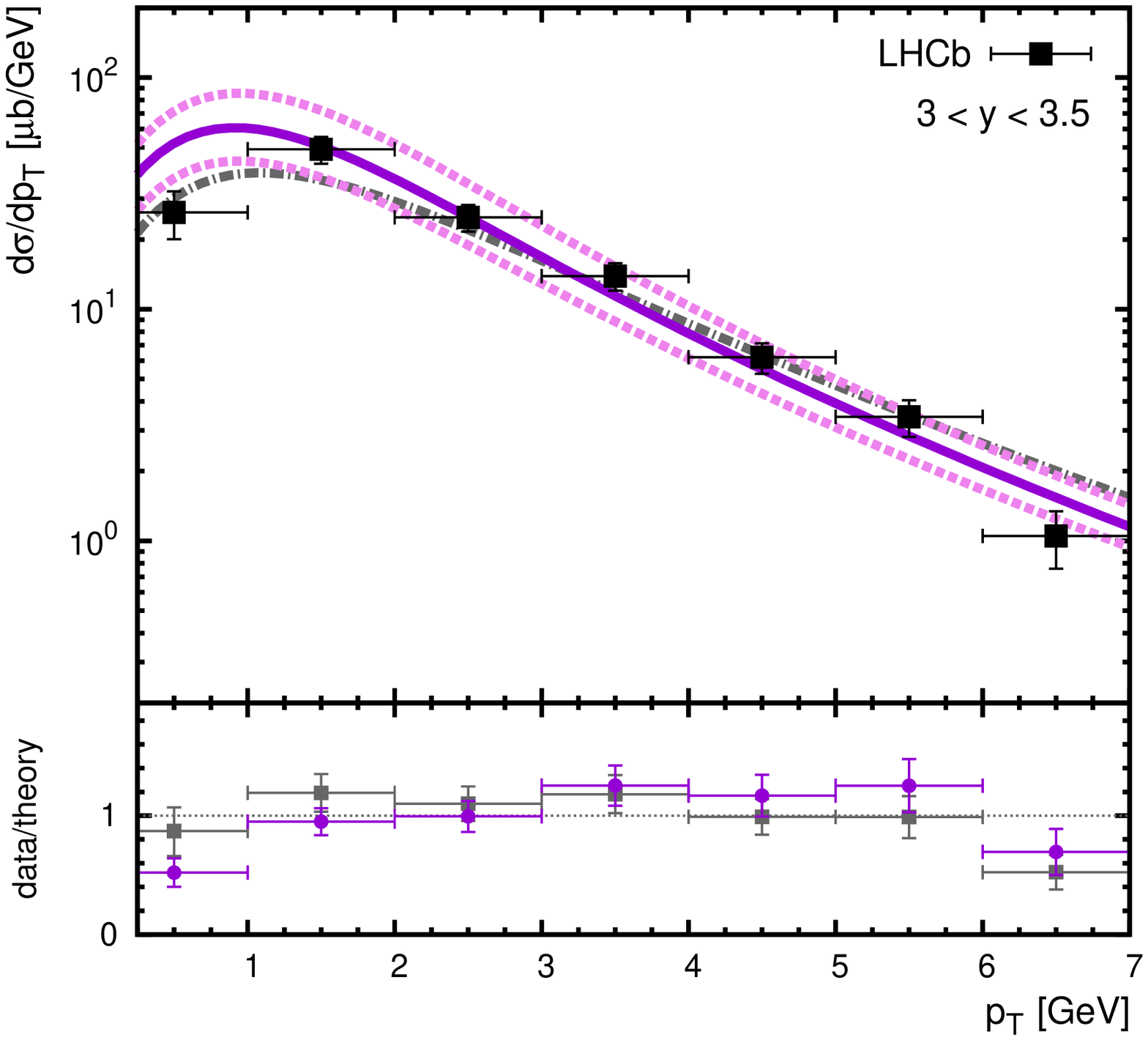, width = 5.5cm, angle = 270}
\hspace{-1cm}
\epsfig{figure=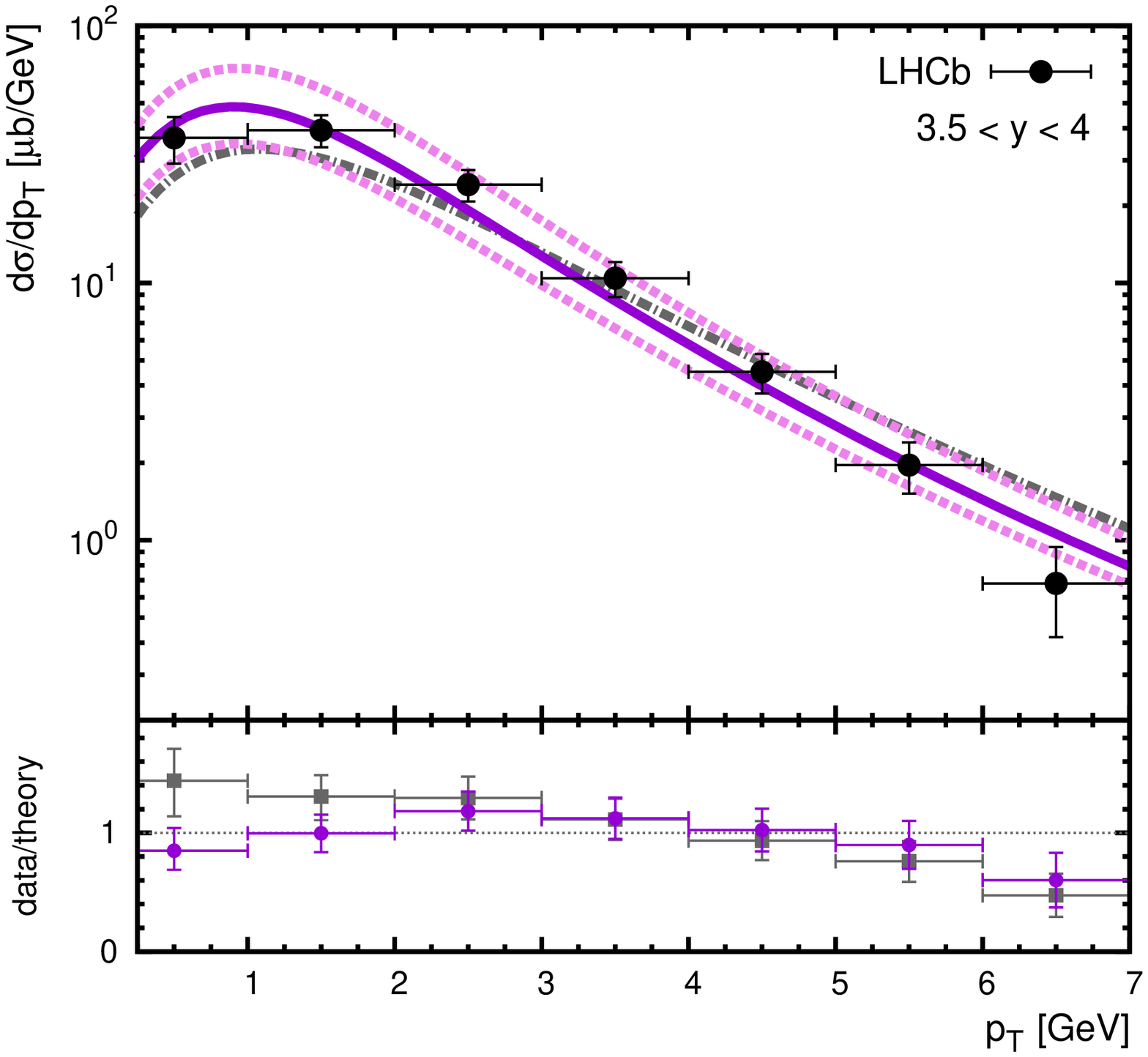, width = 5.5cm, angle = 270}
\epsfig{figure=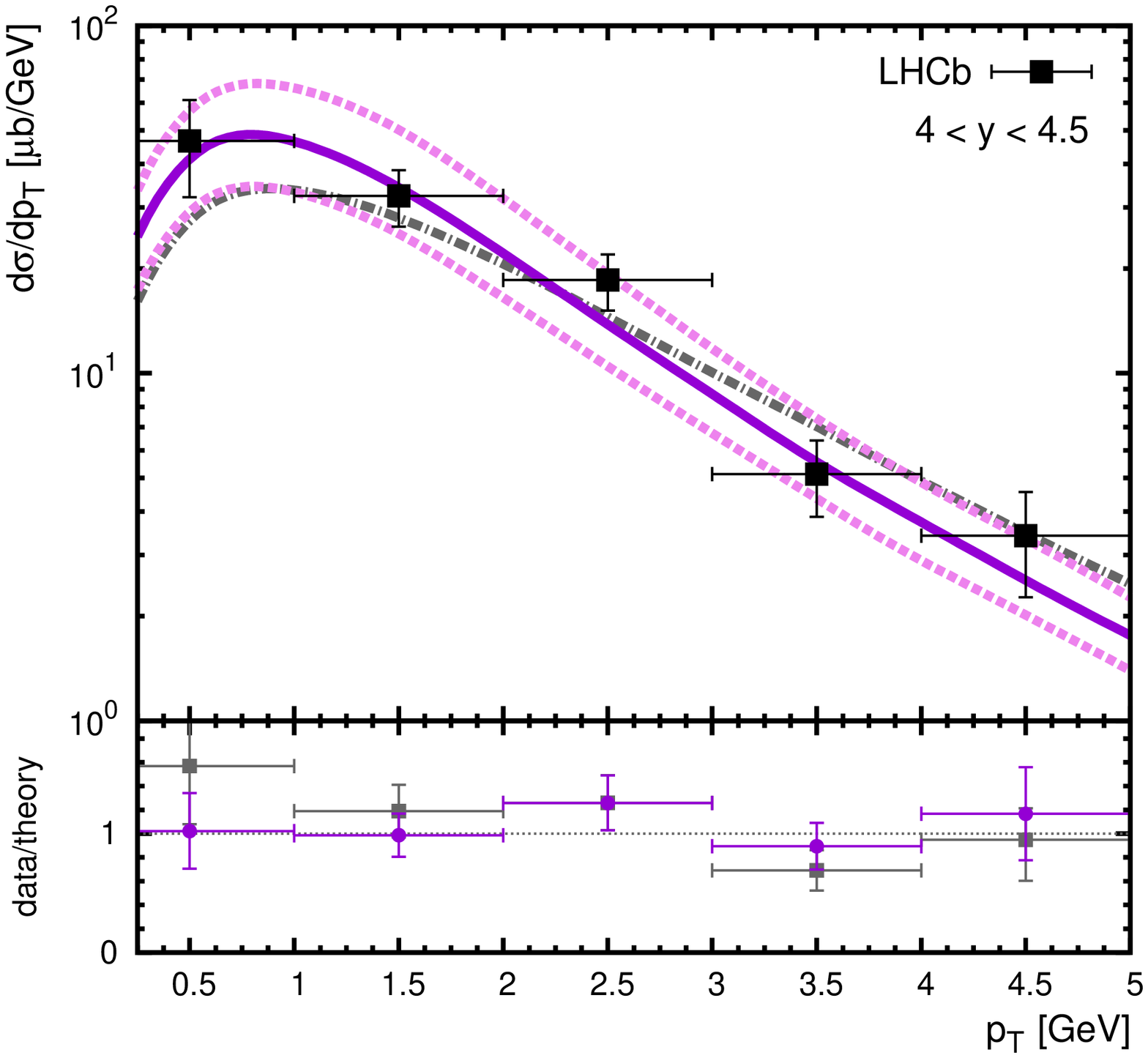, width = 5.5cm, angle = 270}
\caption{The transverse momentum distributions of the $D^*$ meson production at
the LHC. The used kinematical cuts are described in the text. 
Notation of all curves is the same as in Fig.~4. 
The experimental data are from LHCb\cite{52}.}
\label{fig12}
\end{center}
\end{figure}

\begin{figure}
\begin{center}
\epsfig{figure=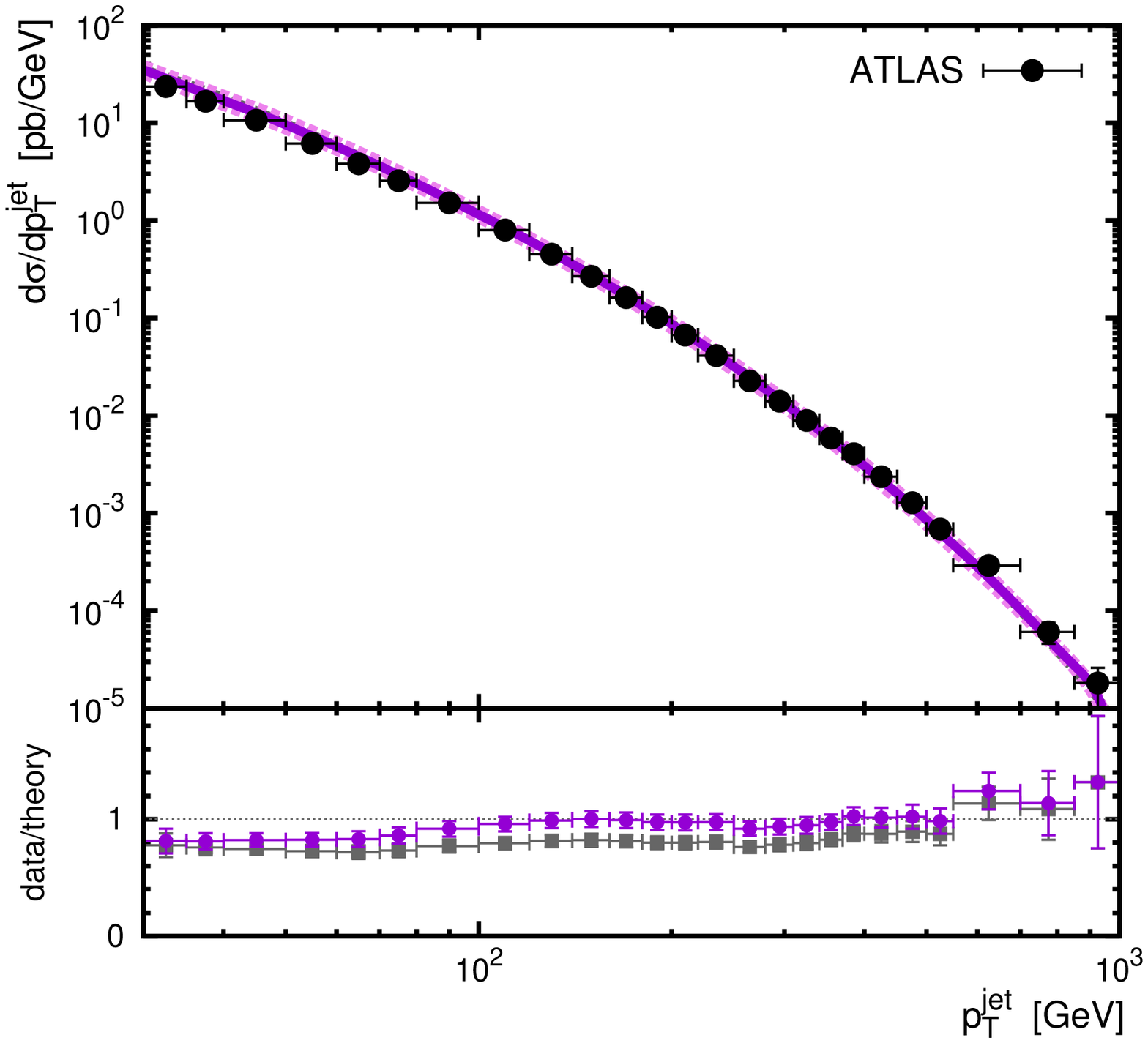, width = 5.5cm, angle = 270}
\vspace{0.7cm} \hspace{-1cm}
\epsfig{figure=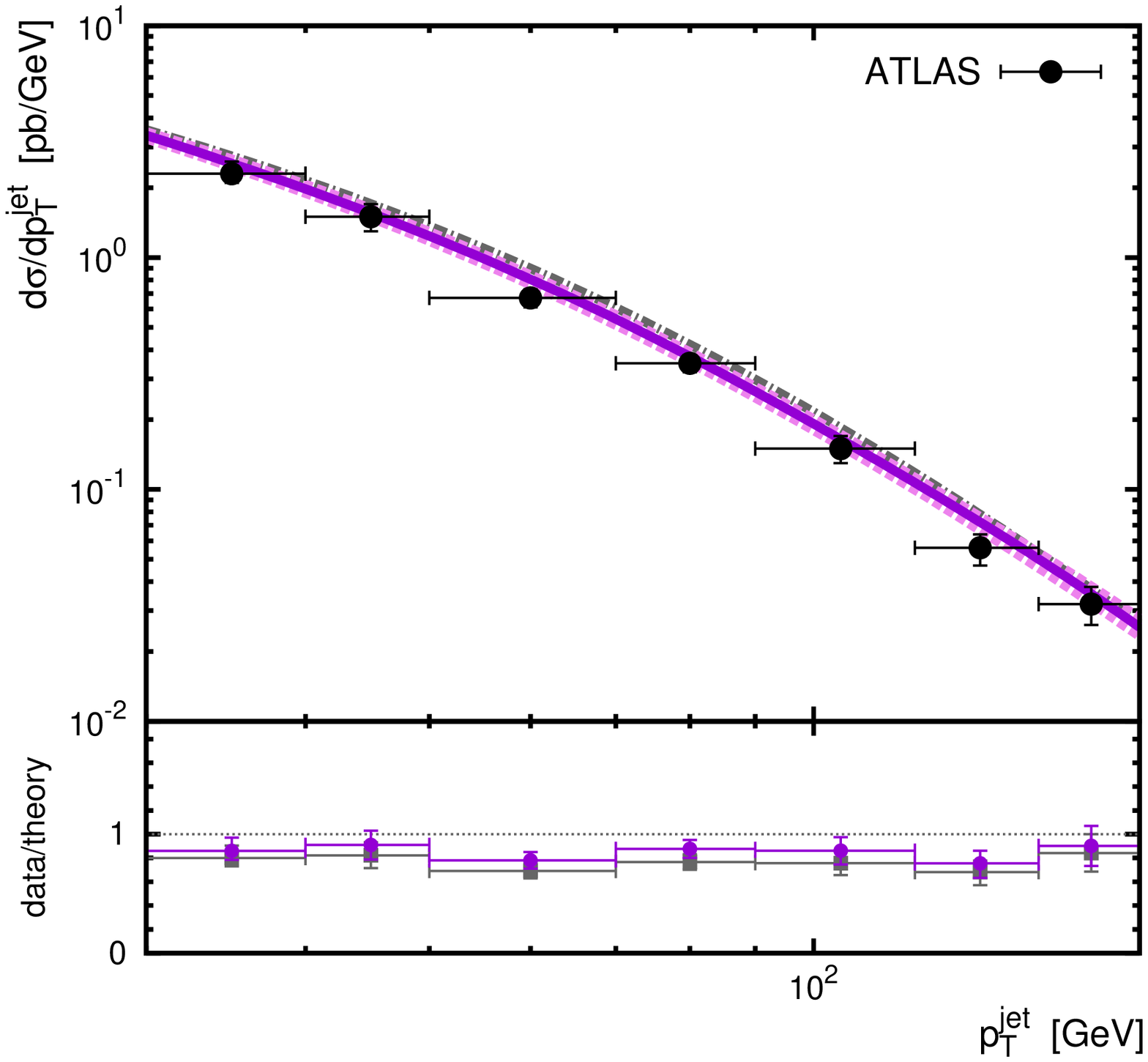, width = 5.5cm, angle = 270}
\caption{The differential cross sections of associated the $W^\pm +$~jet (left panel) 
and $Z/\gamma^* +$~jet (right panel) production 
at the LHC as a function of the jet transverse momentum. 
Notation of all curves is the same as in Fig.~4.
The used kinematical cuts are described in the text. 
The experimental data are from ATLAS\cite{56,57}.}
\label{fig13}
\end{center}
\end{figure}


\begin{thebibliography}{58}
\bibitem{1} L.V.~Gribov, E.M.~Levin, M.G.~Ryskin, Phys. Rep. {\bf 100}, 1 (1983);\\
  E.M.~Levin, M.G.~Ryskin, Yu.M.~Shabelsky, A.G.~Shuvaev, Sov. J. Nucl. Phys. {\bf 53}, 657 (1991).
\bibitem{2} S.~Catani, M.~Ciafaloni, F.~Hautmann, Nucl. Phys. B {\bf 366}, 135 (1991);\\
  J.C.~Collins, R.K.~Ellis, Nucl. Phys. B {\bf 360}, 3 (1991).
\bibitem{3} B.~Andersson {\sl et al.} (Small-$x$ Collaboration), Eur. Phys. J. C {\bf 25}, 77 (2002);\\
  J.~Andersen {\sl et al.} (Small-$x$ Collaboration), Eur. Phys. J. C {\bf 35}, 67 (2004);\\
  J.~Andersen {\sl et al.} (Small-$x$ Collaboration), Eur. Phys. J. C {\bf 48}, 53 (2006).
\bibitem{4} J.C.~Collins, {\it Foundations of perturbative QCD}, Cambridge University Press, 2011.
\bibitem{5} E.~Avsar, arXiv:1108.1181 [hep-ph]; arXiv:1203.1916 [hep-ph].
\bibitem{6} S.M.~Aybat, T.C.~Rogers, Phys. Rev. D {\bf 83}, 114042 (2011).
\bibitem{7} F.~Hautmann, M.~Hentschinski, H.~Jung, Nucl. Phys. B {\bf 865}, 54 (2012). 
\bibitem{8} B.I.~Ermolaev, M.~Greco, S.I.~Troyan, Eur. Phys. J. C {\bf 71}, 1750 (2011);\\
  B.I.~Ermolaev, M.~Greco, S.I.~Troyan, Eur. Phys. J. C {\bf 72}, 1953 (2012).
\bibitem{9} P.~Kotko, K.~Kutak, C.~Marquet, E.~Petreska, S.~Sapeta, A.~van~Hameren, JHEP {\bf 1509}, 106 (2015).
\bibitem{10} A.A.~Grinyuk, A.V.~Lipatov, G.I.~Lykasov, N.P.~Zotov, Phys. Rev. D {\bf 87}, 074017 (2013).
\bibitem{11} A.V.~Lipatov, G.I.~Lykasov, N.P.~Zotov, Phys. Rev. D {\bf 89}, 014001 (2014).
\bibitem{12} K.~Golec-Biernat, M.~W\"usthoff, Phys. Rev. D {\bf 59}, 014017 (1998);\\
  K.~Golec-Biernat, M.~W\"usthoff, Phys. Rev. D {\bf 60}, 114023 (1999).
\bibitem{13} M.~Ciafaloni, Nucl. Phys. B {\bf 296}, 49 (1988);\\
  S.~Catani, F.~Fiorani, G.~Marchesini, Phys. Lett. B {\bf 234}, 339 (1990);\\
  S.~Catani, F.~Fiorani, G.~Marchesini, Nucl. Phys. B {\bf 336}, 18 (1990);\\
  G.~Marchesini, Nucl. Phys. B {\bf 445}, 49 (1995).
\bibitem{14} E.A.~Kuraev, L.N.~Lipatov, V.S.~Fadin, Sov. Phys. JETP {\bf 44}, 443 (1976);\\
  E.A.~Kuraev, L.N.~Lipatov, V.S.~Fadin, Sov. Phys. JETP {\bf 45}, 199 (1977);\\
  I.I.~Balitsky, L.N.~Lipatov, Sov. J. Nucl. Phys. {\bf 28}, 822 (1978).
\bibitem{15} V.N.~Gribov and L.N.~Lipatov, Sov.J. Nucl. Phys. {\bf 15}, 438 (1972);\\
  L.N.~Lipatov, Sov. J. Nucl. Phys. {\bf 20}, 94 (1975);\\
  G.~Altarelli, G.~Parisi, Nucl. Phys. B {\bf 126}, 298 (1977);\\
  Yu.L.~Dokshitzer, Sov. Phys. JETP {\bf 46}, 641 (1977).
\bibitem{16} D.A.~Artemenkov, G.I.~Lykasov, A.I.~Malakhov, Int. J. Mod. Phys. A {\bf 30}, 1550127 (2015); arXiv:1504.07841 [hep-ph].
\bibitem{17} V.A.~Bednyakov, A.A.~Grinyuk, G.I.~Lykasov, M.~Poghosyan, Int. J. Mod. Phys. A {\bf 27}, 1250042 (2012). 
\bibitem{18} Yu.V.~Kovchegov, Phys. Rev. D{\bf 61}, 074018 (2000).
\bibitem{19} A.B.~Kaidalov, Z. Phys. C {\bf 12}, 63 (1982); Surveys High Energy Phys. {\bf 13}, 265 (1999);\\
  A.B.~Kaidalov, O.I.~Piskunova, Z. Phys. C {\bf 30}, 145 (1986); Yad. Fiz. {\bf 43}, 1545 (1986).
\bibitem{20} G.I.~Lykasov, M.N.~Sergeenko, Z. Phys. C {\bf 56} 697, (1992); Z. Phys. C {\bf 52} 635, (1991); Z. Phys. C {\bf 70}, 455 (1996).
\bibitem{21} V.A.~Bednyakov, G.I.~Lykasov, V.V.~Lyubushkin, Europhys. Lett. {\bf 92}, 31001 (2010); arXiv:1005.0559 [hep-ph].
\bibitem{22} NA61 Collaboration, Eur. Phys. J. C {\bf 74}, 2794 (2014).
\bibitem{23} ATLAS Collaboration, New J. Phys. {\bf 13}, 053033 (2011);\\
  CMS Collaboration, Phys. Rev. Lett. {\bf 105}, 022002 (2010).
\bibitem{24} V.S.~Fadin, L.N.~Lipatov, Phys. Lett. B {\bf 429}, 127 (1998).
\bibitem{25} M.~Ciafaloni, G.~Camici, Phys. Lett. B {\bf 430}, 349 (1998).
\bibitem{26} D.N.~Triantafyllopoulos, Nucl. Phys. B {\bf 648}, 293 (2003).
\bibitem{27} M.~Deak, F.~Hautmann, H.~Jung, K.~Kutak, arXiv:1012.6037 [hep-ph].
\bibitem{28} M.~Deak, F.~Hautmann, H.~Jung, K.~Kutak, Eur. Phys. J. C {\bf 72}, 1982 (2012).
\bibitem{29} A.D.~Martin, W.J.~Stirling, R.S.~Thorne, G.~Watt, Eur. Phys. J. C {\bf 63}, 189 (2009).
\bibitem{30} F.~Hautmann, H.~Jung, S.~Taheri Monfared, Eur. Phys. J. C {\bf 74}, 3082 (2014).
\bibitem{31} S.~Catani, F.~Hautmann, Nucl. Phys. B {\bf 427}, 475 (1994); Phys. Lett. B {\bf 315}, 157 (1993).
\bibitem{32} G.~Curci, W.~Furmanski, R.~Petronzio, Nucl. Phys. B {\bf 175}, 27 (1980).
\bibitem{33} F.~Hautmann, M.~Hentschinski, H.~Jung, arXiv:1207.6420 [hep-ph].
\bibitem{34} H.~Jung, arXiv:hep-ph/0411287.
\bibitem{35} A.V.~Kotikov, A.V.~Lipatov, G.~Parente, N.P.~Zotov, Eur. Phys. J. C {\bf 26}, 51 (2002);\\
  A.V.~Kotikov, A.V.~Lipatov, N.P.~Zotov, Eur. Phys. J. C {\bf 27}, 219 (2003). 
\bibitem{36} H.~Jung, M.~Kraemer, A.V. Lipatov, N.P. Zotov, Phys. Rev. D {\bf 85}, 034035 (2012). 
\bibitem{37} N.P.~Zotov, A.V.~Lipatov, V.A.~Saleev, Phys. Atom. Nucl. {\bf 66}, 755 (2003);\\
  S.P.~Baranov, N.P.~Zotov, A.V. Lipatov, Phys. Atom. Nucl. {\bf 67}, 837 (2004). 
\bibitem{38} A.V. Lipatov, N.P. Zotov, Phys. Rev. D {\bf 90}, 094005 (2014);\\ 
A.V.~Lipatov, M.A.~Malyshev, N.P.~Zotov, JHEP {\bf 1112}, 117 (2011). 
\bibitem{39} PDG Collaboration, Chin. Phys. C {\bf 38}, 090001 (2014).
\bibitem{40} G.P.~Lepage, J. Comput. Phys. {\bf 27}, 192 (1978).
\bibitem{41} H1 Collaboration, Eur. Phys. J. C {\bf 74}, 2814 (2014). 
\bibitem{42} ZEUS Collaboration, Phys. Lett. B {\bf 682}, 8 (2009). 
\bibitem{43} ZEUS Collaboration, JHEP {\bf 1409}, 127 (2014). 
\bibitem{44} H1 Collaboration, Eur. Phys. J. C {\bf 71}, 1769 (2011); Eur. Phys. J. C {\bf 72}, 2252 (2012). 
\bibitem{45} H1 Collaboration, Eur. Phys. J. C {\bf 65}, 89 (2010). 
\bibitem{46} CMS Collaboration, JHEP {\bf 1204}, 084 (2012).
\bibitem{47} ATLAS Collaboration, Eur. Phys. J. C {\bf 71}, 1846 (2011). 
\bibitem{48} CMS Collaboration, Phys. Rev. Lett. {\bf 106}, 112001 (2011); JHEP {\bf 1103}, 090 (2011). 
\bibitem{49} ATLAS Collaboration, JHEP {\bf 1310}, 042 (2013). 
\bibitem{50} LHCb Collaboration, JHEP {\bf 1308}, 117 (2013). 
\bibitem{51} C.~Peterson, D.~Schlatter, I.~Schmitt, P.~Zerwas, Phys. Rev. D {\bf 27}, 105 (1983).
\bibitem{52} LHCb Collaboration, Nucl. Phys. B {\bf 871}, 1 (2013). 
\bibitem{53} M.~Cacciari, P.~Nason, Phys. Rev. Lett. {\bf 89}, 122003 (2002);\\
  M.~Cacciari, S.~Frixione, M.L.~Mangano, P.~Nason, G.~Ridolfi, JHEP {\bf 0407}, 033 (2004).
\bibitem{54} M.~Cacciari, P.~Nason, JHEP {\bf 0309}, 006 (2003).
\bibitem{55} E.~Braaten, K.-M.~Cheng, S.~Fleming, T.C.~Yuan, Phys. Rev. D {\bf 51}, 4819 (1995).
\bibitem{56} ATLAS Collaboration, Eur. Phys. J. C {\bf 75}, 82 (2015). 
\bibitem{57} ATLAS Collaboration, Phys. Rev. D {\bf 85}, 032009 (2012). 
\bibitem{58} H.~Jung, Comp. Phys. Comm. {\bf 143}, 100 (2002);\\
  H.~Jung {\sl et al.}, Eur. Phys. J. C {\bf 70}, 1237 (2010).
  
\end{thebibliography}
\end{document}